\documentclass[fleqn,usenatbib]{mnras}
\usepackage{txfonts}

\usepackage[T1]{fontenc}

\DeclareRobustCommand{\VAN}[3]{#2}
\let\VANthebibliography\thebibliography
\def\thebibliography{\DeclareRobustCommand{\VAN}[3]{##3}\VANthebibliography}

\usepackage{graphicx}	
\usepackage{amssymb}	
\usepackage[nointegrals]{wasysym} 
\usepackage{float}
\usepackage{epstopdf}
\usepackage{booktabs}

\DeclareUnicodeCharacter{2212}{-}
\DeclareUnicodeCharacter{0301}{\'{e}}

\title[Time resolved spectroscopy of GRS 1915+105 flare]{Time resolved spectroscopy of a GRS 1915+105 flare during its unusual low state using \emph{AstroSat}}

\author[Sajad et al.]{
Sajad Boked$^{1}$\thanks{E-mail: sboked10@gmail.com},
Bari Maqbool$^{2}$,
V. Jithesh$^{3}$,  
Ranjeev Misra$^{4}$\thanks{E-mail: rmisra@iucaa.in},
Naseer Iqbal$^{1,2}$,  
and Yashpal Bhulla$^{5}$
\\
$^{1}$Department of Physics, University of Kashmir, Srinagar 190006, India\\
$^{2}$Islamic University of Science and Technology Awantipora, Pulwama 192122, India \\
$^{3}$Department of Physics and Electronics, CHRIST (Deemed to be University), Hosur Main Road, Bengaluru - 560029, India\\
$^{4}$Inter-University Centre for Astronomy and Astrophysics, Post Bag 4, Ganeshkhind, Pune 411007, India\\
$^{5}$Pacific Academy of Higher Education and Research University, Udaipur-313003,India
}

\date{Accepted XXX. Received YYY; in original form ZZZ}

\pubyear{2022}

\begin{document}
\label{firstpage}
\pagerange{\pageref{firstpage}--\pageref{lastpage}}
\maketitle

\begin{abstract}
Since its discovery in 1992, GRS 1915+105 has been among the brightest sources in the X-ray sky. However, in early 2018, it dimmed significantly and has stayed in this faint state ever since. We report on \emph{AstroSat} and \emph{NuSTAR} observation of GRS 1915+105 in its unusual low/hard state during 2019 May. We performed time-resolved spectroscopy of the X-ray flares observed in this state and found that the spectra can be fitted well using highly ionised absorption models. We further show that the spectra can also be fitted using a highly relativistic reflection dominated model, where for the lamp post geometry, the X-ray emitting source is always very close to the central black hole. For both interpretations, the flare can be attributed to a change in the intrinsic flux, rather than dramatic variation in the absorption or geometry. These reflection dominated spectra are very similar to the reflection dominated spectra reported for Active Galactic Nuclei in their low flux states.

\end{abstract}




\section{Introduction} \label{introduction}
\quad Black hole (BH) X-ray binary (XRB) system consists of a BH and its companion which is a normal star, orbiting around their common centre of mass. The BH accretes matter from its companion via a geometrically thin, optically thick accretion disc \citep{shakura1973, novikov1973}. Depending on the mass of the companion, those having a less massive companion (M $\leq$ \(\textup{M}_\odot\)) usually accrete via Roche-lobe overflow and are referred as Low Mass X-ray Binaries (LMXBs) while those having a more massive companion (M $>$ \(\textup10{M}_\odot\)) usually accrete through stellar winds and are referred as High Mass X-ray Binaries (HMXBs) \citep{petterson1978, blondin1991,liu2007, Tan_2021}. The majority of BH-XRBs are observed as X-ray transients \citep{corral2016, Tetarenko2016} and most of them are LMXBs. LMXBs spend most of their lives in a quiescent state where there is a steady mass transfer from the companion onto the compact object. An increase in the temperature of the disc is believed to initiate thermal viscous instabilities that prompt a fast outflow of material from the companion onto the black hole resulting in an X-ray outburst. \citep{shakura1976, hameury1990black}. The outburst duration in a BH-XRB can range from several months to a few years. State changes are a common feature in BH-XRBs. During a typical outburst, they evolve in their spectral and timing properties, leading to the classification of various spectral states. \citep{Remillard2006, Belloni2016}. Typically, an outburst begins with a low/hard state, progresses to a high/soft state, and ultimately reverts back to a low/hard state towards the end of the outburst. Intermediate states are often observed around the time of state transitions \citep{Belloni2000}.

Three main radiative components are often present in the X-ray spectrum of BH-XRBs. The black-body radiation from the accretion disc dominating the emission in the soft X-ray band \citep{shakura1973}, power-law like component caused by inverse Compton scattering of the disc photons from the region near the central BH, typically referred to as \emph{corona}, dominating the emission in the hard X-ray band \citep{shakura1976, Lightman1987, sunyaev1980}. The lower and upper cutoffs of this power-law component is determined by the temperature of the seed photons and electrons, respectively. The third component arises due to the fraction of the coronal photons irradiating the disc and being scattered into our line of sight. These photons are reprocessed in the disc's atmosphere and produce reflection spectrum with  characteristic features. The scattering produces a broad Compton hump peaking around $\sim$20–30 keV \citep{fabian1989, lightman1988} and many spectral lines, the strongest being a broad emission line at $\sim$6.4 keV due to iron \citep{george1991, ross2005}. The line profile is set by special and general relativistic effects.

GRS 1915+105 is among one of the most well studied BH-XRBs, discovered in 1992 as an X-ray transient with the \emph{WATCH} instrument onboard International Astrophysical Observatory \emph{“GRANAT”} \citep{1992IAUC.5590....2C}. GRS 1915+105 hosts a BH of mass $12.5^{+2.0}_{-1.8}$\(\textup{M}_\odot\), located at radio parallax distance of $8.6^{+2.0}_{-1.6}$kpc with disc inclination of 60$^{\circ}$ \citep{reid2014parallax}. GRS 1915+105 has the longest orbital period of 33.9 days \citep{2013ApJ...768..185S} among all the LMXBs and therefore has the largest accretion disc and high accretion rate. It is a unique source among all BH-XRBs because of its interesting properties, viz. high luminosity, distinctive X-ray variability classes \citep{Belloni2000}, high frequency quasi-periodic oscillations \citep{Remillard2006} , probably due to large accretion disc and high accretion rate. After a long period of about 26 years in outburst, in July, 2018 ($\sim$MJD 58300) GRS 1915+105 entered into unusually low X-ray flux state. \cite{2018ATel11828....1N} reported the lowest soft X-ray flux of the source in 22 years of continuous monitoring with \emph{MAXI/GSC} and \emph{RXTE/ASM}. The strong similarities between this flux drop and outburst evolution of other BH-XRBs led \cite{2018ATel11828....1N} to conclude that GRS 1915+105 was finally entering into a quiescent state, but in May, 2019 ($\sim$MJD 58600) the source again showed strong activity with a further drop in average X-ray flux and has remained in this faint state since then. Strong radio flares in this state also indicate that the mass accretion rate can still be high \citep{2019ATel12773....1M}. In this obscured state GRS 1915+105 has flux an order of magnitude less than observed previously also strong erratic X-ray flares have been observed in this state \citep{Iwakiri, 2019ATel12793....1N,jitesh}. This kind of variability has not been seen in this source before. X-ray observations in this state showed Compton thick obscuration, hard spectra, with strong emission and absorption lines characterised by a clear increase in intrinsic absorption with column densities over an order of magnitude larger than those usually observed in the source before. \citep{motta2021observations, 2019ATel12771....1M,2019ATel12788....1M, 2021ApJ...909...41B, miller2020obscured, koljonen2020obscured}.

Reflected spectra characterised by high reflection fraction (exceeding unity) have been observed in many individual observations of low flux states in Active Galactic Nuclei (AGN) (Seyferts) MCG-6-30-15 \citep{fabian2003iron}, 1H 0707-495 \citep{2004MNRAS.353.1071F} and 1H0419-577 \citep{fabian2005x}. This high reflection fraction was explained on the basis of light bending model as discussed by \citep{2004MNRAS.349.1435M}. In their model X-ray source is assumed to be very close to the maximally rotating central BH, so a large number of photons will be captured by the central BH and most of the remaining photons will be focused towards the innermost regions of the accretion disc. The strong light bending effects in the innermost parts of the accretion disc enhance the reflection fraction thus producing the reflection dominated spectra. \cite{2004MNRAS.349.1435M} in their light bending model identified three different flux states i.e. low, intermediate and high in which the reflection dominated component of the spectrum is  correlated, anti-correlated or almost independent with respect to the direct continuum. 
 
Highly reflection dominated spectra have been observed in some BH-XRBs also. \cite{rossi2005iron, 2013ApJ...763...48R} analysed the \emph{Rossi X-ray Timing Explorer (RXTE)} data of XTE J1650-500 and showed that the reflection fraction increases sharply at transition between soft intermediate and hard intermediate states. They also showed that at low power-law flux levels the Fe K$\alpha$ line flux and power-law flux are positively correlated and line equivalent width and power-law flux are anti-correlated which is in agreement with the light bending model proposed by \cite{2004MNRAS.349.1435M}.  

\cite{2020A&A...639A..13K} analysed \emph{NuSTAR} spectra for a number of black hole X-ray binaries (V4641 Sgr, Cyg X-3 , V404 Cyg and GRS 1915+105) using a complex combination of spectral components such as smeared edges, partial covering, distant and relativistic reflection and obscuration/emission from surrounding tori. In general, their preference is that these sources exhibit either highly absorbed partial covering or can be explained in terms of a torus covering the source. In particular, for GRS 1915+105, they analysed three epochs and found that the latter two epochs require highly absorbed partial covering with column density $\sim 5 \times 10^{23} cm^{-2}$ or a torus with radial velocity. The first epoch could be represented by their torus model, but also by relativistic reflection.

The \emph{AstroSat} observation used in this study was taken just a day after \emph{MAXI/GSC} detected the re-brightening of GRS 1915+105 \citep{Iwakiri}. The bottom panel of the \emph{MAXI} lightcurve shown in Figure~\ref{fig:maxi} indicates that the source count rate was almost three times large during the \emph{AstroSat} observation period (depicted by solid lines) compared to the \emph{NuSTAR} observation period (depicted by dashed lines). The strong erratic flares were observed in both the \emph{LAXPC} and \emph{SXT} lightcurves, where the source intensity increased by a factor of $\sim$5. The flaring activity became weak after first 60 ksec time of the lightcurve and then disappeared completely. The detailed spectral study of this observation was not done before. In this work we report the results of our analysis of one of the flares for which simultaneous \emph{LAXPC} and \emph{SXT} data is available. The detailed analysis of the other flares will be discussed in follow-up work. One of the \emph{NuSTAR} observations (\cite{2020A&A...639A..13K} Epoch 1 data) has spectrum similar to the one observed by \emph{AstroSat} and hence we re-analyse the data using the models used for \emph{AstroSat}. The structure of this paper is as follows. In Section~\ref{data_reduction} the observations and data reduction procedures are described. The findings from spectral analysis of our time-resolved spectra are discussed in Section~\ref{spectral}. These results are then discussed and interpreted in Section~\ref{discussion}. We then discuss our conclusions in Section~\ref{Conclusions}.

\section{Observation and data reduction} \label{data_reduction}

\begin{figure*}
	\includegraphics[scale=0.5]{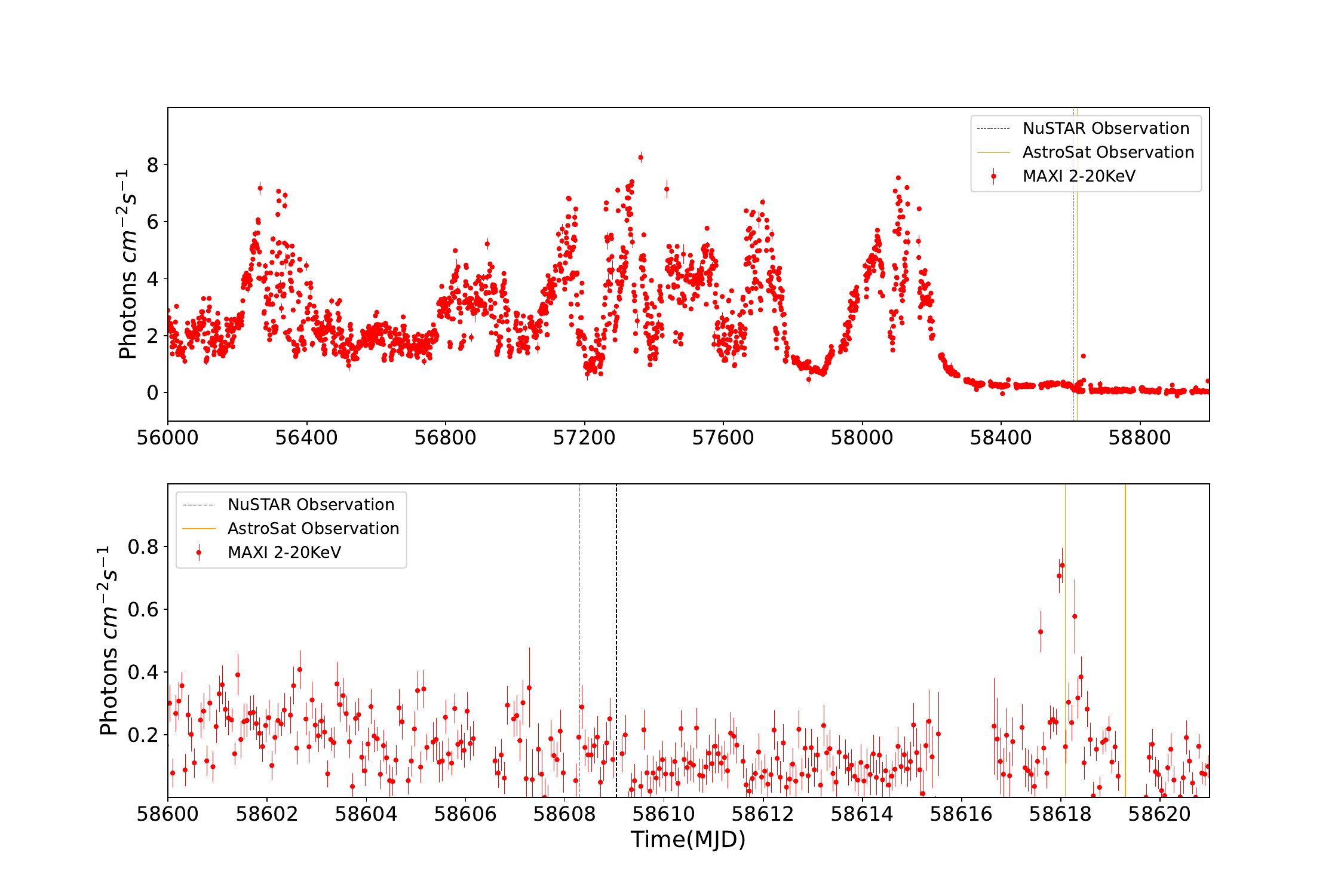}
    \caption{Upper panel: One-day average \emph{MAXI/GSC} long term lightcurve of GRS 1915+105 from MJD 56000 to 59500 (2012-2022). The dashed and solid solid lines corresponds to the time of our \emph{NuSTAR} and \emph{AstroSat} observations shown in Table~\ref{obslog}. Lower panel: One orbit binned \emph{MAXI/GSC} lightcurve from MJD 58600 to 58620 showing the time of our \emph{NuSTAR} and \emph{AstroSat} observation marked between dashed and solid vertical lines respectively.}
    \label{fig:maxi}
\end{figure*}

In this work, we used \emph{AstroSat} and \emph{NuSTAR} observations of GRS 1915+105 conducted during May 2019. The observation details are given in Table ~\ref{obslog}. These observations are also overplotted on the one-day averaged  2--20 keV \emph{MAXI} light curve shown in Figure~\ref{fig:maxi}. The one orbit binned \emph{MAXI} light curve showing the observation time of both the \emph{NuSTAR} and \emph{AstroSat} observations is also shown in Figure ~\ref{fig:maxi}.

\begin{table*}
\centering
\caption{Observation log.}
\label{obslog}
\begin{tabular}{lccccccc}
\hline
Instrument &Obs ID & Start time & End time & MJD  &  Exposure \\
& & Date &Date &   & (ks) \\
\hline
{\it AstroSat} & T03\_116T01\_9000002916 & 02:04:22 & 07:14:37 & 58618.09064  & 22.6 \\
 				&						& 15/05/2019 & 16/05/2019 & \\
{\it NuSTAR} & 90501321002 & 07:06:09 &01:01:09  & 58608.29974  & 28.7 \\
  				&						& 05/05/2019 & 06/05/2019 & \\

\hline
\end{tabular}
\end{table*}

\subsection{NuSTAR} \label{nustar_reduction}
\begin{figure}
	
	\includegraphics[scale=0.65]{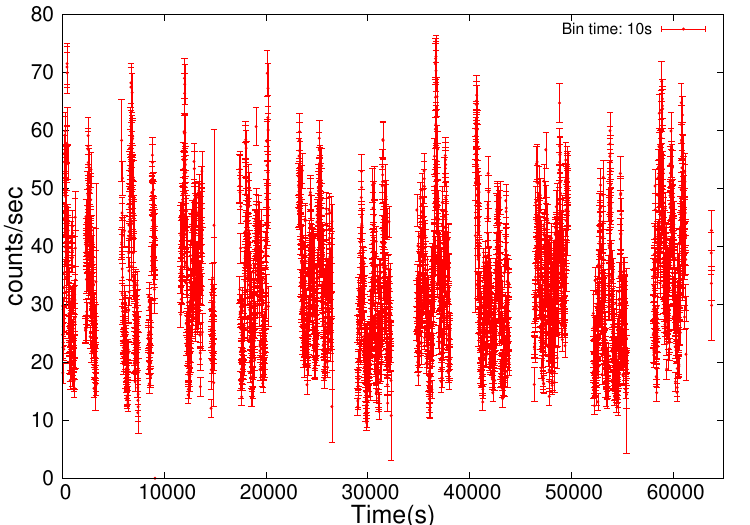}
    \caption{\emph{NuSTAR} lightcurve of the observation used for this study. The time along X-axis is the time since 58608.29974 MJD (start time of observation). }
    \label{fig:nustar}
\end{figure}

The Nuclear Spectroscopic Telescope Array \emph{(NuSTAR)} consists of two co-aligned, identical X-ray telescopes focusing the hard X-rays in a wide energy range of 3–79 keV. Each of the telescopes has its own focal plane modules \emph{FPMA} and \emph{FPMB} \citep{harrison2013nuclear}.
We selected the \emph{NuSTAR} observation of GRS 1915+105 during 2019 May 5-6 (Observation ID: 90501321002) in its unusual low state. We reduced the raw data following the \emph{NuSTAR} Data Analysis Software Guide \footnote{\url{https://heasarc.gsfc.nasa.gov/docs/nustar/analysis/nustar_swguide.pdf}} using the \emph{NuSTAR} data analysis software ({\sc NuSTARDAS v2.1.1}) provided under \emph{HEASOFT V6.29} with CALDB version (20221229). The filtered clean event files were produced  using the {\sc nupipeline} routine by setting the parameters "TENTACLE=YES" and "SAAMODE=OPTIMISED" to remove times with high background and the science products for both \emph{FPMA and FPMB} were then extracted using the {\sc nuproducts} routine. The source spectrum was extracted by taking a circular region of radius 100\arcsec centred on the source and the background spectrum from the circular region of radius 100\arcsec of the detector not contaminated by source counts. The data from both the detectors \emph{FPMA and FPMB} is grouped with background, ARF and RMF using {\sc grppha} command. An optimal binning scheme using {\sc ftgrouppha} was then used to group the energy bins together. The light curve of the \emph{NuSTAR} observation is shown in Figure ~\ref{fig:nustar}.

\subsection{AstroSat} \label{astrosat_reduction}
\emph{AstroSat} is India's first multi-wavelength astronomical satellite dedicated to observing cosmic sources in wide energy ranges of the electromagnetic spectrum, from optical to X-ray bands \citep{2006ESASP.604..907A}. \emph{AstroSat} has five different payloads with  Soft X-ray Telescope \emph{(SXT)} \citep{2016SPIE.9905E..1ES, singh2017soft} and Large Area X-ray Proportional Counter \emph{(LAXPC)} \citep{yadav2016astrosat, 2017ApJS..231...10A}) providing an opportunity to study the X-rays in energy range of 0.3--8.0 keV and 3.0--80.0 keV respectively.

\begin{figure}
	
	\includegraphics[scale=0.47]{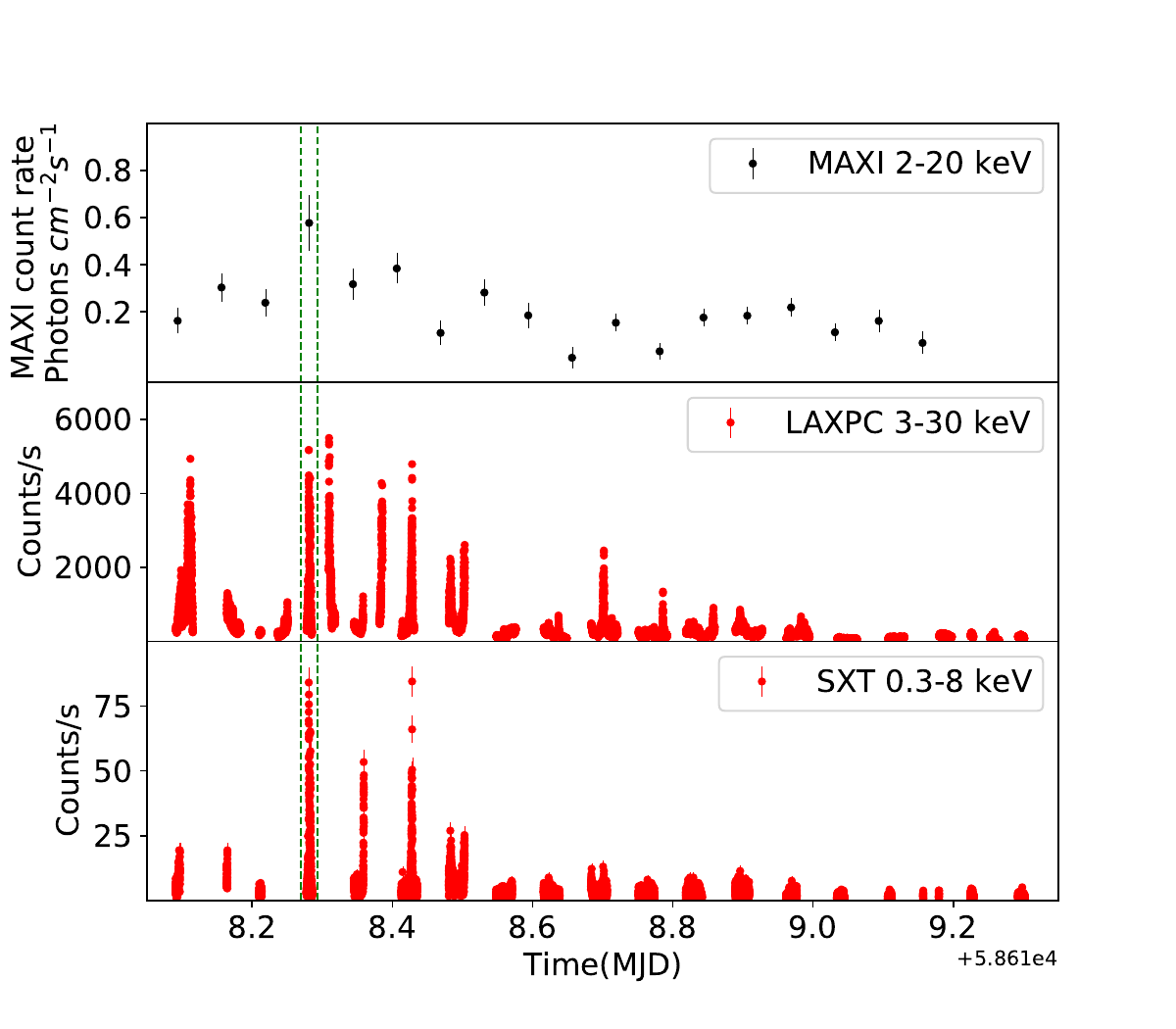}
    \caption{\emph{MAXI} and \emph{Astrosat} lightcurves:top panel shows zoomed 1.5 hour binned lightcurve in energy range 2.0--20.0 keV for the time of \emph{AstroSat} observation, the middle panel corresponds to background subtracted \emph{LAXPC} lightcurve in 3.0--30.0 keV energy band and bottom panel corresponds to \emph{SXT} lightcurve in 0.3--8.0 keV energy band. The flare marked between dotted vertical lines was only used for this analysis.}
    \label{fig:astrosat}
\end{figure}
\begin{figure}
	
	\includegraphics[scale=0.39]{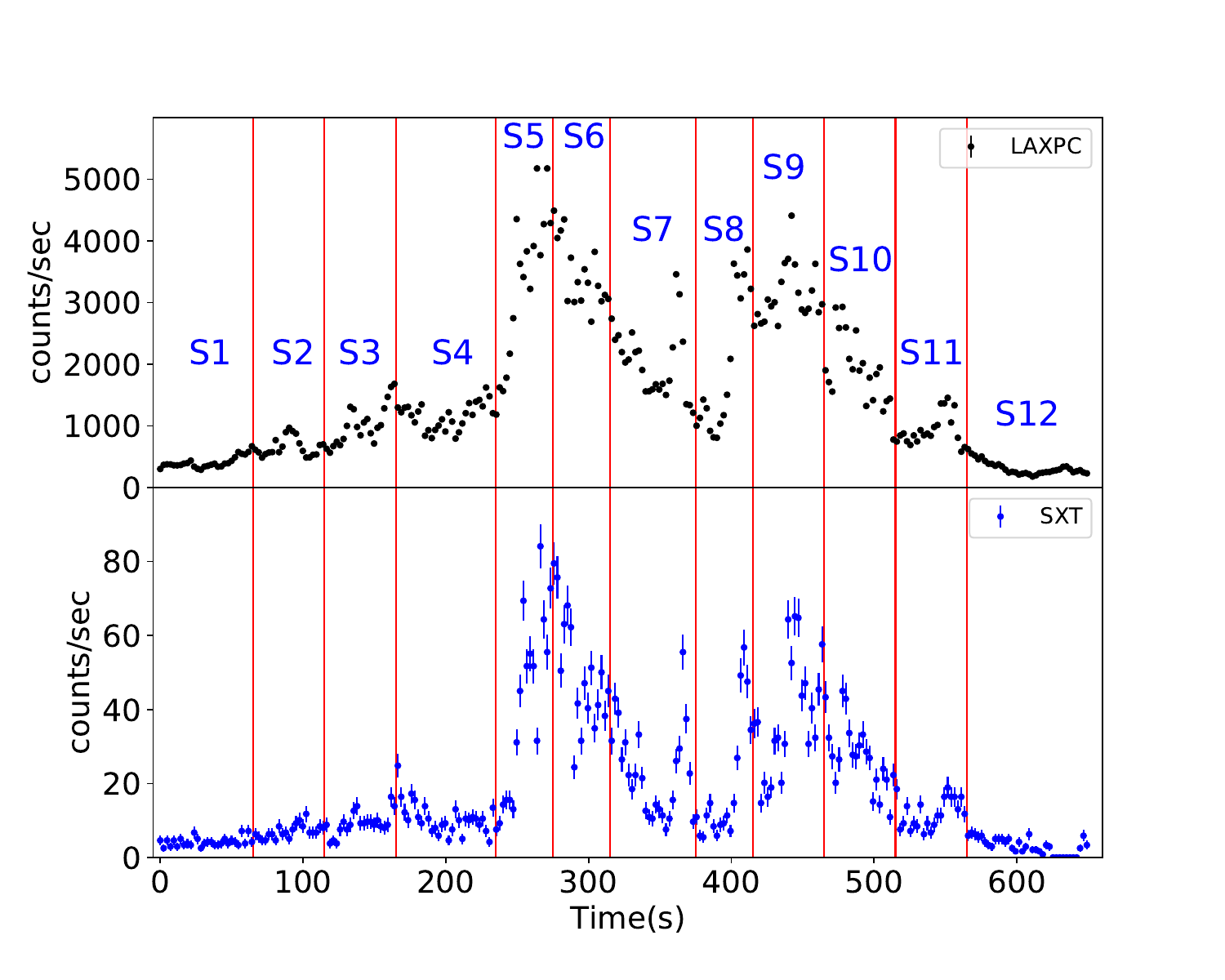}
    \caption{Zoomed in view of the flare marked in Figure~\ref{fig:astrosat} considered for this study. The time along X-axis is the time since 295598385 \emph{AstroSat} seconds (start of flare). The flare is divided into twelve different segments from S1 to S12 and are marked by the vertical lines. Upper panel is for \emph{LAXPC} while the lower panel shows the \emph{SXT} lightcurve. }
    \label{fig:flare4}
\end{figure}

The SXT level 2 data for 16 orbits in photon counting (PC) mode was downloaded from \textit{AstroSat data archive-ISSDC}. The data reduction was achieved with tools provided by \emph{SXT POC} team at TIFR \footnote{\url{https://www.tifr.res.in/~astrosat_sxt/dataanalysis.html}}. The individual clean event files for each orbit were then combined using \textit{SXTPYJULIAMERGER\_v01} to generate a single merged clean event file. To extract the science products, further analysis of data was performed using \emph{XSELECT V2.4m}. We extracted a circular source region of radius 18 arc min centred on the source. For segments S5,S6 and S9, since the net count rate was $>$40 counts $s^{-1}$, to account for pileup effect we extracted annulus source region of inner radius 2.5 arcmin and outer radius of 18 arc mins. We have used response matrix file(RMF) \footnote{sxt\_pc\_mat\_g0to4.rmf} and blank sky background spectrum file \footnote{SkyBkg\_comb\_EL3p5\_Cl\_Rd16p0\_v01.pha} provided by the \emph{SXT} instrument team for our analysis. The vignetting corrected ancillary response file (ARF) was created with \textit{sxtARFModule} using the ARF file provided by \emph{SXT} team. The background spectrum, vignetting corrected ARF and RMF were then grouped together using interactive command \textit{grppha} such that each bin contains a minimum of 25 counts. 

\emph{LAXPC} consists of three proportional counters \emph{LAXPC10, LAXPC20 and LAXPC30}. We have used only \emph{LAXPC20} data for our analysis as \emph{LAXPC10} and \emph{LAXPC30} were not functional at the time of our observation. We downloaded the level 1 data for all the 16 orbits from the \emph{AstroSat} data archive and used the \emph{LAXPC} software (August 15, 2022 version) \footnote{\url{http://astrosat-ssc.iucaa.in/laxpcData}} to convert it to level 2. \textit{laxpc\_make\_lightcurve} and \textit{laxpc\_make\_spectra} codes were then used to extract the lightcurve and spectrum fot each segment.

The simultaneous \emph{LAXPC} and \emph{SXT} lightcurves of the observation along with \emph{MAXI} lightcurve of the source for the time of \emph{AstroSat} observation is shown in Figure ~\ref{fig:astrosat}. The \emph{LAXPC} and \emph{SXT} lightcurves are 2.3775 sec binned, the minimum time resolution of \emph{SXT} instrument. Among the erratic X-ray flares in \emph{LAXPC} and \emph{SXT} lightcurves, we have selected the flare marked between dotted lines in Figure ~\ref{fig:astrosat} for our study. We chose this specific flare because it occurred simultaneously in both the \emph{LAXPC} and \emph{SXT} instruments, and had both the highest count rate and longer flaring duration among all other flares having simultaneous data available for both the instruments. We then divided the flare into twelve different segments, as shown in Figure ~\ref{fig:flare4} and both \emph{LAXPC} and \emph{SXT} products were extracted for each segment using single gti for both instruments.

\section{SPECTRAL ANALYSIS} \label{spectral}
All the spectral fits were performed in XSPEC version: 12.12.0 \citep{1996ASPC..101...17A}. A constant factor was introduced in all the joint spectral fits to account for the cross-calibration uncertainty between different instruments. For \emph{AstroSat} analysis the constant was fixed at unity for \emph{SXT} and was allowed to vary for \emph{LAXPC}, while for \emph{NuSTAR} constant was fixed at unity for \emph{FPMA} and was allowed to vary for \emph{FPMB}. In addition to this a gain fit was also performed on both the \emph{LAXPC} and \emph{SXT} to modify the gain of the response file \citep{article}. A systematic error of 3\% was used in the joint fitting of \emph{AstroSat SXT} and \emph{LAXPC} spectra to take care of the uncertainties in the response matrix. The error on all the parameters are reported at 90\% confidence level unless specified otherwise.

The \emph{NuSTAR} data used in this study was analysed by \cite{2020A&A...639A..13K} and they reported the presence of strong iron line feature, a strong iron absorption line and curvature in the hard X-rays around $\sim$20--30 keV. This curvature being the typical feature for hard state XRB might be caused by heavy absorption of the incident spectrum or by Compton down-scattering in the accretion disc or in the surrounding medium. For the \emph{NuSTAR} observation, \cite{2020A&A...639A..13K} used \emph{relxillCp}, a gaussian line to model the strong iron absorption line around
 around 6.5 keV, the interstellar absorption was taken care of using \textit{phabs}. A smeared iron absorption edge component \textit{smedge} was also used to model the absorption features in the spectra around 7--8 keV. 
 
 \subsection{Spectral fits with absorption dominated model} \label{abs}
\begin{figure*}
    \centering
     \includegraphics[scale=0.29,angle=-90]{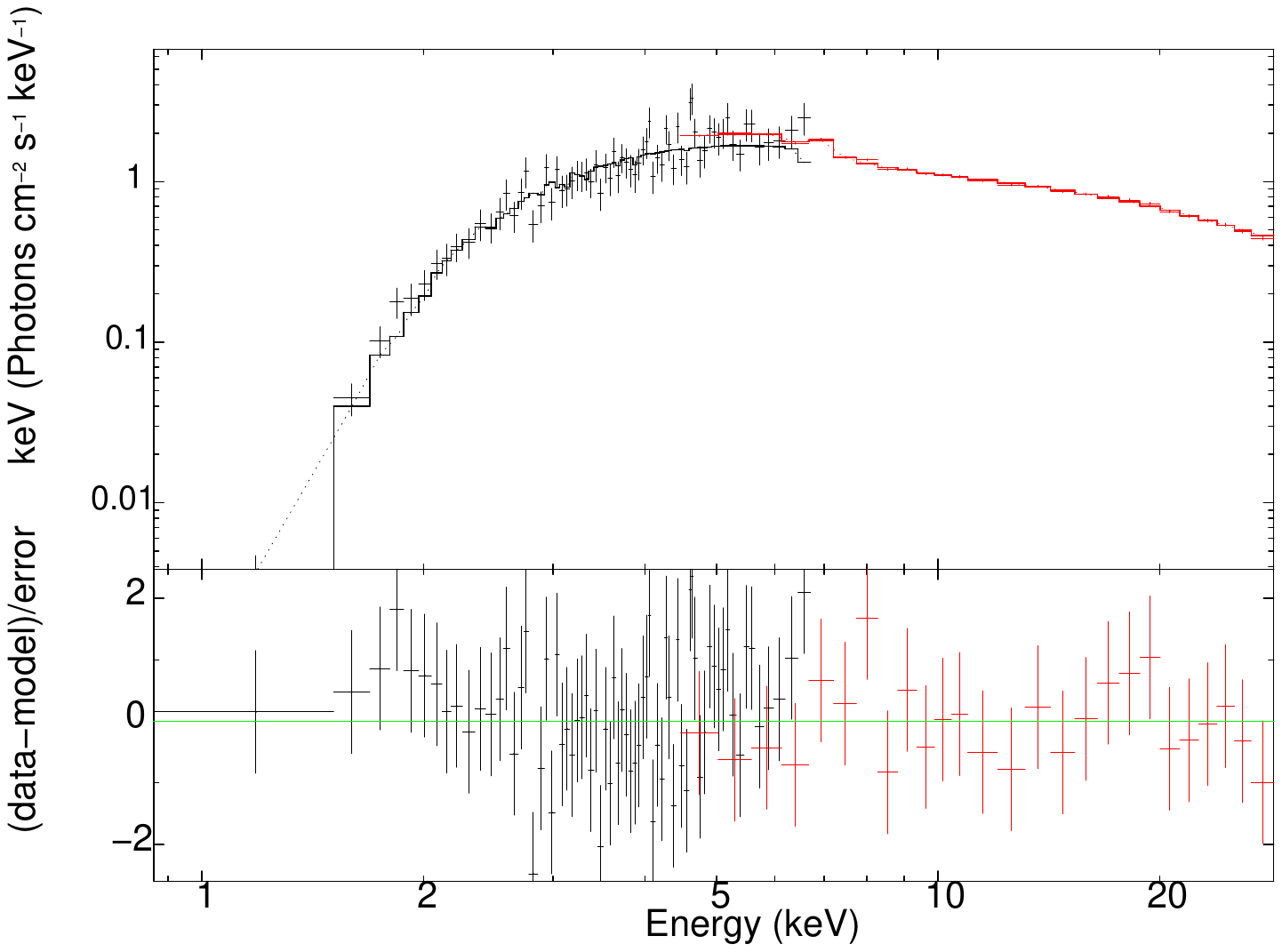}
    \includegraphics[scale=0.29,angle=-90]{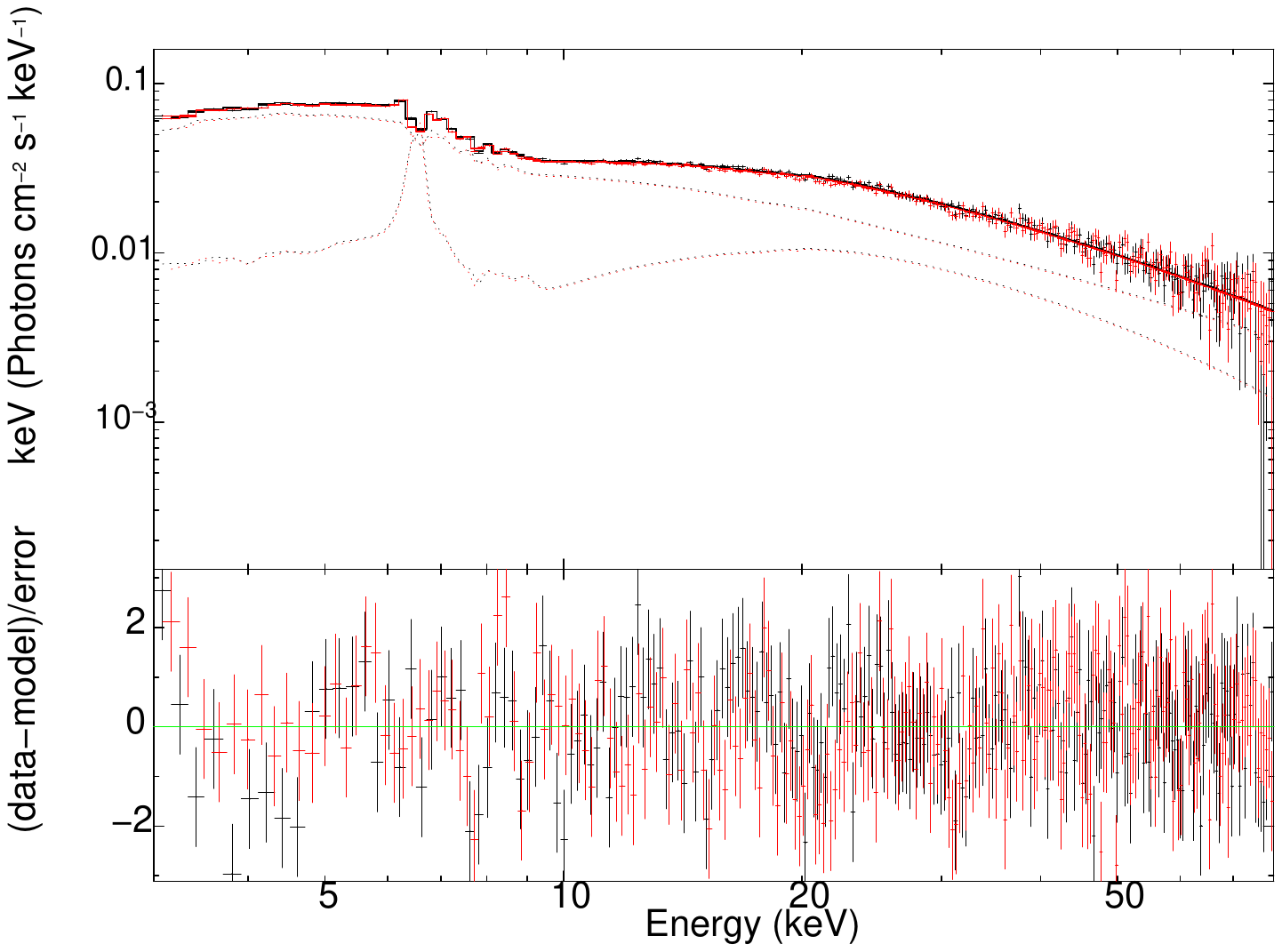}
    \caption{The left panel represents the \emph{AstroSat} spectrum of segment 6 from the simultaneous fit of 0.7--7 keV \emph{SXT} and 4--30 keV \emph{LAXPC} data modelled with {\sc constant$\times$tbabs$\times$zxipcf(thcomp$\times$diskbb + xillvercp)} and the right panel shows the \emph{NuSTAR FPMA} and \emph{FPMB} spectrum in energy range 3--79 keV modelled using {\sc constant$\times$tbabs$\times$zxipcf(thcomp$\times$diskbb + xillvercp +gaussian)}.}
    \label{fig:zxipcf_plots}
\end{figure*}

The spectra for all the segments from \emph{Astrosat} and \emph{NuSTAR} were modelled using a Comptonisation component \textit{thcomp}, a multicoloured black-body component \textit{diskbb} and a reflection component \textit{xillverCp}, to model the distant reflection from the disc. The Galactic absorption was taken care of by the absorption model \emph{tbabs} and the local obscured absorption features observed in the spectrum was modelled using a photo-ionised absorption model \textit{zxipcf}\citep{Reeves_2008} . In addition to this, a \textit{gaussian} line was used to model the strong iron absorption line at $\sim$6.5 keV for the \emph{NuSTAR} data. So the total model used is {\sc constant$\times$tbabs$\times$zxipcf(thcomp$\times$diskbb + xillverCp)} in XSPEC notation for the \emph{AstroSat} data and an additional \textit{gaussian} component for the \emph{NuSTAR} data. The neutral hydrogen column density was fixed at 5.0$\times10^{22}cm^{-2}$. This is in comparison to the earlier observed values \cite{zoghbi,koljonen2020obscured}. Initially \textit{zxipcf} covering fraction was left free to vary for \emph{AstroSat} data, we could not bound its upper limit and its value was always consistent with unity, so we fixed it at 1, i.e. we assumed the total covering of the X-ray source, while for the \emph{NuSTAR} data it was left free to vary. The inner disc temperature of the disc i.e the parameter \emph{Tin} in model \textit{diskbb} was not constrained by the data, and hence was fixed at 1 keV. \textit{xillverCp} was included only as a reflection component by fixing the reflection fraction at -1. The iron abundance $A_{FE}$, in units of solar abundance was fixed at 1. The density of the disc in logarithmic units (logN) and the inclination of the disc were fixed at 17 $cm^{-3}$ and 60$^{\circ}$. All the parameters of the Comptonisation component describing the coronal properties were tied to the corresponding \textit{xillverCp} values. The electron temperature (kTe) for the \emph{AstroSat} data was fixed at 400 keV, since its value was high for \emph{NuSTAR} data and the upper limit was pegged at 400 keV. Adding a systematic error of 1\% in the \emph{NuSTAR} data improved the fit by $\Delta \chi^2 \sim$138 so we have added 1\% systematic error to the spectral fitting of \emph{NuSTAR} data. The systematic also includes complex spectral model systematic rising due to the model assumptions regarding geometry and homogeneity, which may not accurately describe the physical source. For \emph{NuSTAR} data, we have included 1\% systematic to take into account such spectral model uncertainties. We calculated the unabsorbed flux in energy range 0.7--30 keV and 3--79 keV for \emph{AstroSat} and \emph{NuSTAR} data respectively, and the corresponding luminosities were then calculated. The best fit spectral parameters for the \emph{AstroSat} and \emph{NuSTAR} observation are presented in Table~\ref{absorption} and Table~\ref{absorption2}. The fitted spectra for one segment of \emph{AstroSat} data and the \emph{NuSTAR} observation are shown in Figure~\ref{fig:zxipcf_plots}. The column density for the ionised absorption ranged from $10^{23}$ to $10^{24} cm^{-2}$.

\begin{table*}
\centering
\caption{Best fit spectral parameters for the model combination {\sc constant$\times$tbabs$\times$zxipcf(thcomp$\times$diskbb + xillvercp)} for \textit{AstroSat} data 0.7--30.0 keV.}
\begin{tabular}{clllllllllll}
\hline
Segment & zxipcf &zxipcf & diskbb& xillverCp & xillverCp &xillverCp   & $\chi{^2}/$dof & Flux & Luminosity \\
		& N$_{\rm{H}}$ & log~$\xi$ &  norm& $\Gamma$  & log~$\xi$ & norm &  & $\times10^{-9}$ & $\times10^{37}$ \\
		& & &  &  &  &$\times10^{-3}$ & &  &\\
\hline

1& $ 87.74^{ -57.70}_{+ 14.09}$& $ 3.07^{ -0.08}_{+ 0.1}$& $ <31.10$& $ 1.88^{ -0.07}_{+ 0.06}$& $ >3.56$& $ 22.40^{ -4.61}_{+ 4.97}$&36.39/31 & $ 9.59^{ -0.15}_{+ 0.15}$& $ 8.49^{ -0.13}_{+ 0.13}$\\ \\
2& $ 52.11^{ -17.51}_{+ 14.42}$& $ 2.99^{ -0.15}_{+ 0.13}$& $ <49.61$& $ 1.90^{ -0.04}_{+ 0.04}$& $ >4.35$& $ 37.11^{ -7.29}_{+ 6.95}$&42.46/36 & $ 15.38^{ -0.22}_{+ 0.23}$& $ 13.62^{ -0.20}_{+ 0.20}$\\ \\
3& $ 35.51^{ -8.44}_{+ 11.14}$& $ 2.86^{ -0.11}_{+ 0.14}$& $ <70.33$& $ 1.94^{ -0.03}_{+ 0.03}$& $ 3.96^{ -0.10}_{+ 0.30}$& $ 58.11^{ -9.77}_{+ 8.32}$&35.20/42 & $ 21.78^{ -0.29}_{+ 0.29}$& $ 19.28^{ -0.25}_{+ 0.26}$\\ \\
4& $ 29.90^{ -5.74}_{+ 4.55}$& $ 2.40^{ -0.15}_{+ 0.13}$& $ <136.85$& $ 1.90^{ -0.05}_{+ 0.05}$& $ 3.99^{ -0.15}_{+ 0.19}$& $ 82.55^{ -13.45}_{+ 19.60}$&65.95/51 & $ 31.61^{ -0.47}_{+ 0.59}$& $ 27.99^{ -0.42}_{+ 0.52}$\\ \\
5& $ 12.08^{ -3.39}_{+ 3.39}$& $ 2.54^{ -0.2}_{+ 0.18}$& $ <242.52$& $ 1.89^{ -0.02}_{+ 0.03}$& $ 4.12^{ -0.16}_{+ 0.24}$& $ 144.27^{ -27.82}_{+ 20.52}$&80.92/82 & $ 55.96^{ -0.61}_{+ 0.63}$& $ 49.55^{ -0.54}_{+ 0.55}$\\ \\
6& $ 12.72^{ -2.85}_{+ 3.22}$& $ 2.31^{ -0.13}_{+ 0.13}$& $ <199.51$& $ 1.93^{ -0.03}_{+ 0.03}$& $ 4.12^{ -0.16}_{+ 0.17}$& $ 188.59^{ -26.56}_{+ 23.59}$&84.75/90 & $ 72.66^{ -0.77}_{+ 0.79}$& $ 64.33^{ -0.68}_{+ 0.70}$\\ \\
7& $ 16.14^{ -3.78}_{+ 3.74}$& $ 2.40^{ -0.14}_{+ 0.15}$& $ <173.92$& $ 1.87^{ -0.03}_{+ 0.03}$& $ 3.82^{ -0.10}_{+ 0.10}$& $ 133.01^{ -21.93}_{+ 16.77}$&89.32/80 & $ 47.26^{ -0.51}_{+ 0.52}$& $ 41.85^{ -0.46}_{+ 0.46}$\\ \\
8& $ 26.07^{ -6.24}_{+ 3.95}$& $ 2.43^{ -0.10}_{+ 0.12}$& $ <452.25$& $ 1.92^{ -0.04}_{+ 0.04}$& $ 3.95^{ -0.12}_{+ 0.19}$& $ 148.10^{ -45.27}_{+ 24.65}$&47.57/59 & $ 55.48^{ -0.66}_{+ 0.66}$& $ 49.12^{ -0.58}_{+ 0.59}$\\ \\
9& $ 13.35^{ -2.81}_{+ 2.99}$& $ 2.24^{ -0.11}_{+ 0.12}$& $ <155.20$& $ 1.89^{ -0.03}_{+ 0.03}$& $ 4.08^{ -0.13}_{+ 0.18}$& $ 162.29^{ -22.17}_{+ 20.41}$&100.14/89 & $ 62.28^{ -0.65}_{+ 0.67}$& $ 55.14^{ -0.58}_{+ 0.60}$\\ \\
10& $ 13.28^{ -3.20}_{+ 3.51}$& $ 2.36^{ -0.13}_{+ 0.14}$& $ <142.98$& $ 1.97^{ -0.02}_{+ 0.02}$& $ 3.82^{ -0.08}_{+ 0.08}$& $ 142.10^{ -17.88}_{+ 20.54}$&83.37/82 & $ 51.87^{ -0.59}_{+ 0.59}$& $ 45.92^{ -0.51}_{+ 0.53}$\\ \\
11& $ 29.92^{ -13.45}_{+ 9.74}$& $ 2.87^{ -0.12}_{+ 0.11}$& $ <94.18$& $ 1.89^{ -0.03}_{+ 0.04}$& $ 3.65^{ -0.11}_{+ 0.14}$& $ 70.59^{ -11.42}_{+ 10.79}$&38.57/47 & $ 23.86^{ -0.31}_{+ 0.32}$& $ 21.13^{ -0.28}_{+ 0.28}$\\ \\
12& $ 17.59^{ -10.06}_{+ 11.10}$& $ 2.34^{ -0.35}_{+ 0.25}$& $ 76.73^{ -35.57}_{+ 46.92}$& $ 1.95^{ -0.13}_{+ 0.10}$& $ 2.95^{ -0.15}_{+ 0.41}$& $ 31.01^{ -8.45}_{+ 12.72}$&27.67/29 & $ 9.62^{ -0.16}_{+ 0.17}$& $ 8.52^{ -0.14}_{+ 0.15}$\\

\hline
\end{tabular}
\begin{flushleft}
\small{All the column densities are measured in $10^{22}$ atoms cm$^{-2}$. The Flux reported here is the unabsorbed flux in the energy range 3--79 keV. All the fluxes are measured in $erg {cm}^{-2}{s}^{-1}$ and all the luminosities in $erg {s}^{-1}$. The ionisation parameter is measured in erg cm $s^{-1}$.}
\end{flushleft}
\label{absorption}
\end{table*}

\begin{table}
\centering
\caption{Best fit spectral parameters for the model combination {\sc constant$\times$tbabs$\times$zxipcf(thcomp$\times$diskbb + xillvercp + gaussian)} for \emph{NuSTAR} data 3--79 keV. The Flux reported here is the unabsorbed flux in the energy range 3--79 keV, measured in $erg {cm}^{-2}{s}^{-1}$ and the luminosity in $erg {s}^{-1}$. The ionisation parameter is measured in erg cm $s^{-1}$. }

\label{absorption2}
\begin{tabular}{lccccccccc}
\hline

\multicolumn{3}{l}{\textsc{zxipcf}} \\
\hline
\hspace{0.1cm} N$_{\mathrm{H}}$ & 10$^{22}$ cm$^{-2}$ & $ 64.42^{ -8.62}_{+ 4.38}$\\
\hspace{0.1cm} log $\xi$ && $3.09^{ -0.06}_{+ 0.03} $ \\
\hspace{0.1cm} $f_{cov}$ && $0.73^{ -0.05}_{+ 0.06} $ \\
\hline
\multicolumn{3}{l}{\textsc{diskbb}} \\
\hline
\hspace{0.1cm} norm & & $68.67^{ -1.97}_{+ 3.8} $  \\ 
\hline
\multicolumn{3}{l}{\textsc{xillvercp}} \\
\hline
\hspace{0.1cm} $\Gamma$ && $2.05^{ -0.01}_{+ 0.01}$ \\
\hspace{0.1cm} kTe & &$>345.86$ \\
\hspace{0.1cm} log $\xi$ && $ 2.78^{ -0.03}_{+ 0.02}$ \\
\hspace{0.1cm} norm & $\times$10$^{-3}$  & $5.64^{ -0.58}_{+ 0.59}$  \\ 
\hline
\multicolumn{3}{l}{\textsc{gauss}} \\
\hline
\hspace{0.1cm} E & keV & $ 6.57^{ -0.02}_{+ 0.02}$ \\
\hspace{0.1cm}  $\sigma$& &$ <0.09 $ \\
\hspace{0.1cm} norm & $\times$10$^{-4}$ & $-17.21^{ -3.33}_{+ 3.56}$  \\
\hline
\hspace{0.1cm} Flux & $\times$10$^{-9}$ & $2.91^{ -0.01}_{+ 0.01} $  \\
\hline
\hspace{0.1cm} Luminosity & $\times$10$^{37}$ & $2.59^{- 0.01}_{+ 0.01} $  \\
\hline
\hspace{0.1cm} $\chi^{2}$/d.o.f & & 521.23/464 \\
\hline
\end{tabular}
\begin{flushleft}

\end{flushleft}
\end{table}

\subsection{Spectral fits with reflection dominated model} \label{refl}
\begin{figure*}
    \centering
     \includegraphics[scale=0.29,angle=-90]{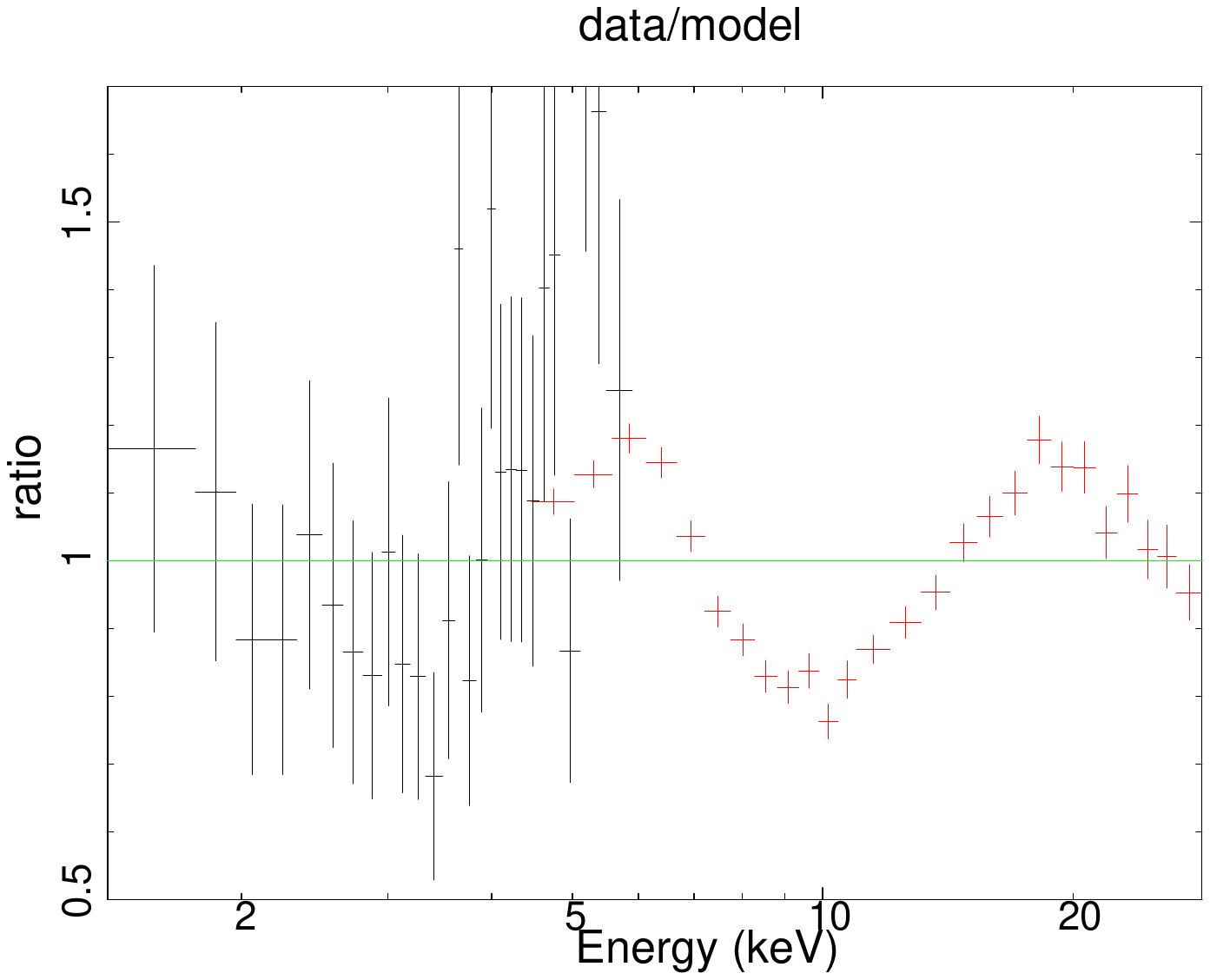}
     \includegraphics[scale=0.29,angle=-90]{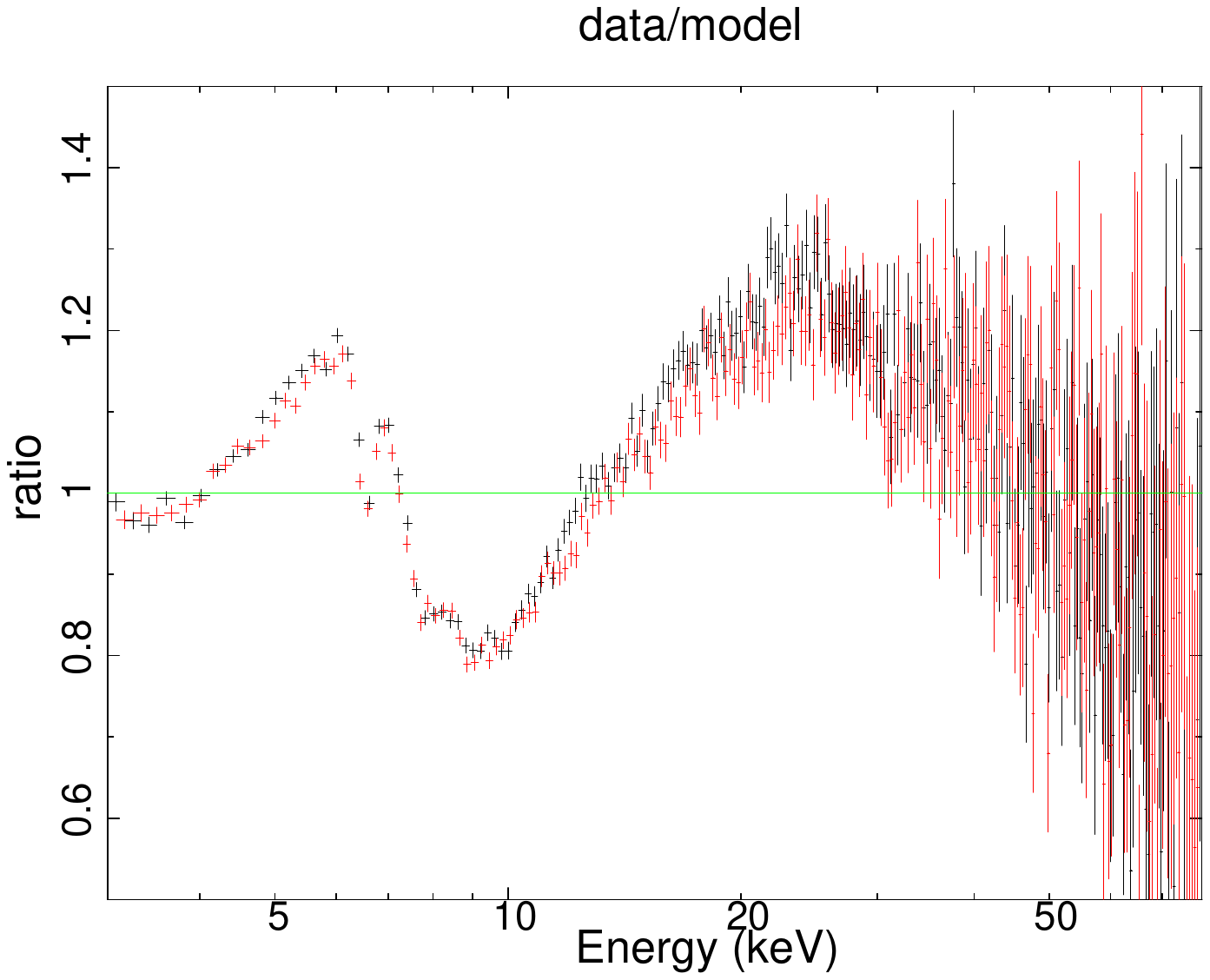}
    \caption{The Ratio plots of \emph{AstroSat} spectrum for segment 11 (Left) and \emph{NuSTAR} spectrum (Right) using model {\sc constant$\times$tbabs$\times$cutoffpl}.}
    \label{fig:ratio_plots}
\end{figure*}

To highlight the reflection features in our spectra we modelled both the \emph{AstroSat} and \emph{NuSTAR} data using the galactic absorption model \emph{Tbabs}, a cutoff power-law \emph{(cutoffpl)} and a constant. We fitted the \emph{LAXPC} and \emph{SXT} spectra in 0.7--30 keV for all the segments and \emph{NuSTAR} spectrum in 3--79 keV. The ratio plots for one of the segments of \emph{AstroSat} data and the \emph{NuSTAR} data are shown in Figure~\ref{fig:ratio_plots}. For both spectra, excess flux is seen around$ \sim 6.5$ and $\sim 20$ keV, indicating the presence of broadened reflection components.

Like the typical hard state spectra of BH-XRBs, we first tried to fit the spectrum with a Comptonisation component \textit{thcomp} \citep{2020MNRAS.492.5234Z}, a multicoloured black-body component \textit{diskbb} \citep{1984PASJ...36..741M} modified by the galactic absorption model \textit{Tbabs} \citep{2000ApJ...542..914W}. There were strong residues in 6.4 keV energy range in the spectra of all the segments, due to the presence of Fe k$\alpha$ emission line and was taken care of using \textit{gaussian}. A constant was also added to the spectra, so our total model in XSPEC notation is {\sc constant$\times$tbabs(thcomp$\times$diskbb + gaussian)}. We fitted all the segments of our \emph{AstroSat} data with this model and it provided good statistical fit to all the spectra with $\chi ^{2}_{reduced}$ comparable to 1 for all segments. However, the electron temperature of the Comptonising component turned out to be $\sim$5 keV with high optical depth $\tau > 12$, making the component peak at $\sim$20 keV. Thus,  the Comptonised component seemed to be modelling the broad reflection component and hence may not be physical. 

\begin{figure*}
    \centering
     \includegraphics[scale=0.29,angle=-90]{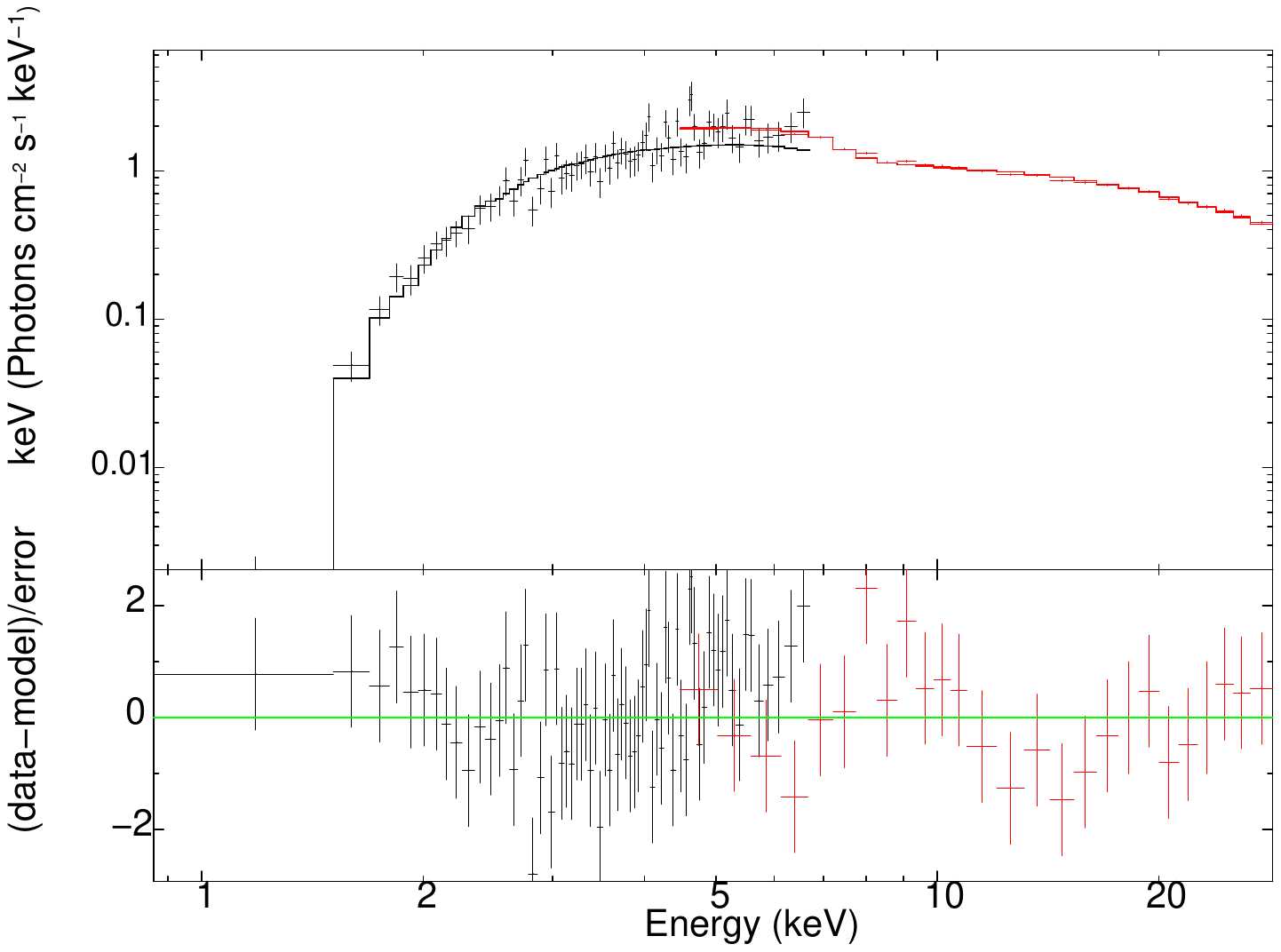}
    \includegraphics[scale=0.29,angle=-90]{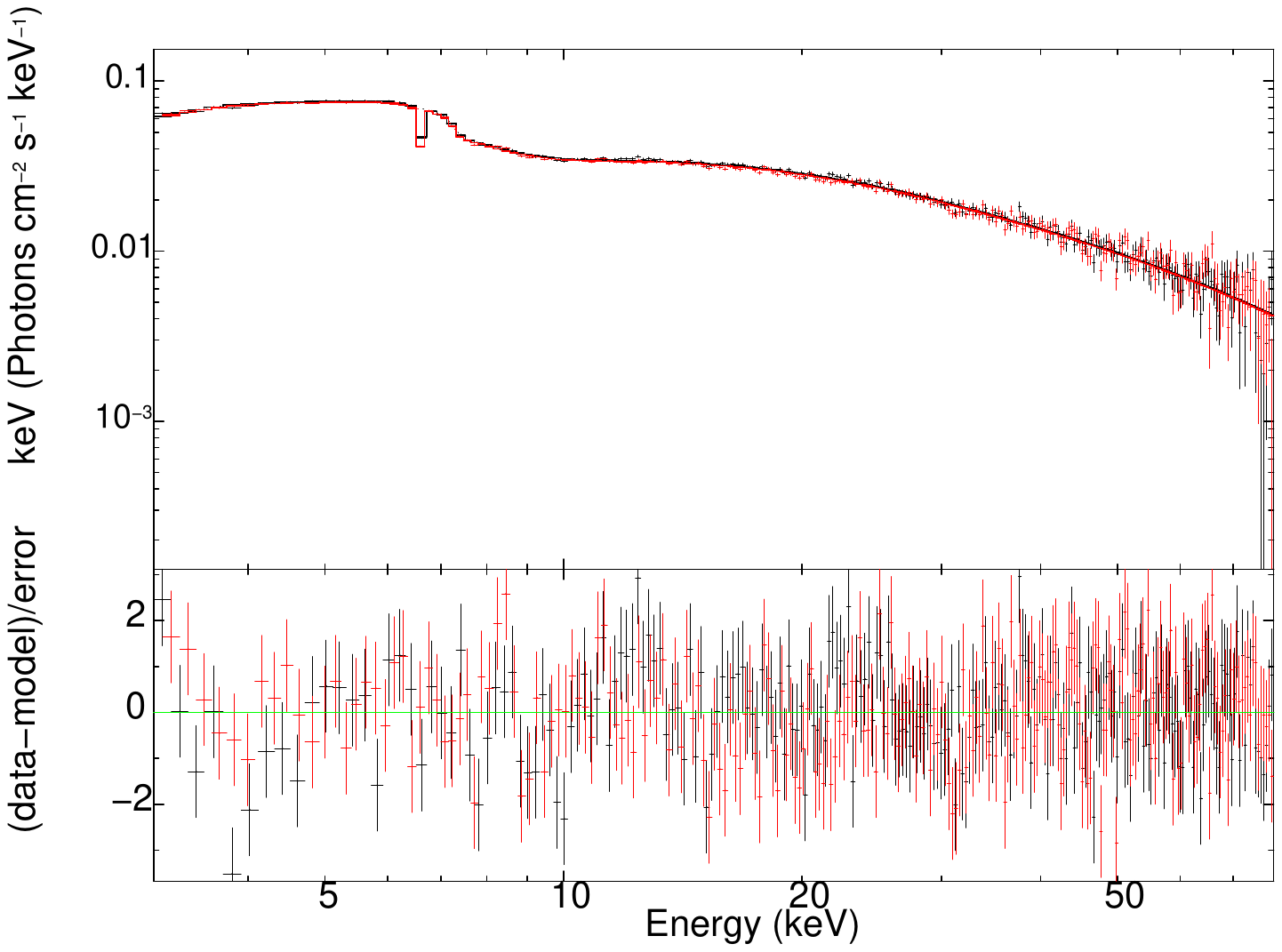}
    \caption{The left panel represents the \emph{AstroSat} spectrum of segment 6 from the simultaneous fit of 0.7--7 keV \emph{SXT} and 4--30 keV \emph{LAXPC} data modelled with {\sc constant$\times$tbabs$\times$relxillcp} and the right panel shows the \emph{NuSTAR FPMA} and \emph{FPMB} spectrum in energy range 3--79 keV modelled using {\sc constant$\times$smedge$\times$tbabs(relxillcp+gaussian)}.}
    \label{fig:relxill_plots}
\end{figure*}

We fitted the joint spectra extracted from \emph{NuSTAR FPMA} and \emph{FPMB} in energy range 3--79 keV and for joint spectra of \emph{SXT} and \emph{LAXPC} the energy range considered was 0.7--7 keV and 4--30 keV. We have used the model \textit{Tbabs} to incorporate the Galactic absorption combined with relativistic reflection model \textit{relxill v 2.2.0} \citep{2014ApJ...782...76G,2014MNRAS.444L.100D} to fit both \emph{AstroSat} and \emph{NuSTAR} data. We used the \textit{relxillCp} flavour of the \textit{relxill} model, which internally includes the original continuum emission modelled using \textit{nthcomp} and also the reflected emission from the disc modified for the relativistic effects in the inner accretion disc near the central BH. Thus the total model used in XSPEC notation is {\sc constant$\times$tbabs$\times$relxillcp}.    While fitting the \emph{NuSTAR} observation with this model we found strong residuals in the energy range (6--8) keV. We found a strong iron absorption line at around 6.5 keV, so we modelled that using \textit{gaussian} and a smeared iron absorption edge \textit{smedge} with index for photo-electric cross-section fixed at -2.67. So the total model we have used for the \emph{NuSTAR} data is {\sc constant$\times$smedge$\times$tbabs(relxillCp + gaussian)} in the XSPEC notation. The constant factor used for the cross-calibration between different instruments was well comparable to 1 for both \emph{NuSTAR} and \emph{AstroSat} observations. Initially, we tried to constraint the reflection fraction, but its value found from the fit was always very high and we thus could not constraint the upper bound on the parameter. We then fitted only the reflection component in \emph{relxillCp} by fixing the model parameter "refl$\_$frac" to negative value of -1. The spin is fixed to 0.998 (see also \citet{grsspin}) to avoid model degeneracies, and this does not provide any indication of the actual value of the intrinsic angular momentum  of the BH. The disc inclination of the system is fixed at 60$^{\circ}$ \citep{reid2014parallax}. The inner radius of the accretion disc $R_{in}$ was kept free to vary and was always found close to the innermost stable circular orbit (ISCO) and the outer radius $R_{out}$ was fixed at $400R_g$, where $R_g$ = $GM/c^2$ is the gravitational radius of the BH. The emissivity profile in the \emph{relxillCp} model is given by $\epsilon(r) \propto$ $r^{-q_{in}}$ for $r<R_{br}$ and $\epsilon(r) \propto$ $r^{-q_{out}}$ for $r>R_{br}$ where $R_{br}$ is the break radius and $q_{in}$ and $q_{out}$, the index for inner and outer regions of the disc. We assumed the single power-law emissivity profile with $q_{in}=q_{out}$ and free to vary. Initially, the density of the accretion disc (logN) was free to vary, and we found the average of the best-fit value from all the segments of AstroSat observation $\sim$17$cm^{-3}$. So we fixed logN to 17$cm^{-3}$. The iron abundance $A_{FE}$ of the source was fixed to solar abundances. For reflection spectra, the \emph{relxill} model allows the ionisation parameter ($\xi$=0) neutral to ($\xi$=4.7) heavily ionised in logarithmic units, where $\xi = 4\pi F/n$, F is the irradiating flux and n is the density of the disc. The electron temperature of the corona (kTe) was very high and its upper limit could not be constrained with our \emph{NuSTAR} data, and the value found was always large than 300 (keV), so we fixed it to 400 (keV) for all the segments. Adding a systematic error of 1\% in the \emph{NuSTAR} data improved the fit by $\Delta \chi^2 \sim$141, so we have added 1\% systematic error to the spectral fitting of \emph{NuSTAR} data. We calculated the flux using XSPEC model \textit{cflux} and the Luminosity was then calculated assuming the distance of the source $8.6^{+2.0}_{-1.6}$kpc. The best fit spectral parameters for the \emph{AstroSat} and \emph{NuSTAR} observation are presented in Table~\ref{reflection} and Table~\ref{reflection2}.The fitted spectra for one segment of \emph{AstroSat} data and the \emph{NuSTAR} observation are shown in Figure~\ref{fig:relxill_plots}. The parameters from the spectral fit of both absorption dominated and reflection dominated models are plotted as a function of time in Figure \ref{fig:par_zxipcf} and \ref{fig:par_relxill}.

\begin{table*}
\centering
\caption{Best fit spectral parameters for the model combination {\sc constant$\times$tbabs$\times$relxillCp} for \textit{AstroSat} data 0.3--30.0 keV.}
\begin{tabular}{clllllllllllllll}
\hline
Segment & Tbabs & relxillCp & relxillCp & relxillCp & relxillCp & relxillCp & $\chi{^2}/$dof & Flux & Luminosity \\
		& N$_{\rm{H}}$ & $R_{in}$ & Emissivity & $\Gamma$  &log~$\xi$ & norm &  & $\times10^{-9}$ & $\times10^{37}$ \\
		& &(ISCO) & Index & & & $\times10^{-3}$ & & & &\\
\hline
1& $ 6.19^{ -0.93}_{+ 0.87}$& $ 1.63^{ -0.13}_{+ 0.08}$& $ >6.92$& $ 1.99^{ -0.06}_{+ 0.08}$& $ 2.80^{ -0.07}_{+ 0.11}$& $ 49.28^{ -7.41}_{+ 10.07}$&32.74/31 & $ 4.74^{ -0.52}_{+ 0.53}$& $ 4.20^{ -0.46}_{+ 0.47}$\\ \\
2& $ 6.95^{ -0.91}_{+ 1.17}$& $ 1.32^{ -0.13}_{+ 0.15}$& $ 6.80^{ -0.98}_{+ 2.14}$& $ 2.12^{ -0.13}_{+ 0.08}$& $ 2.78^{ -0.12}_{+ 0.16}$& $ 102.18^{ -29.27}_{+ 28.70}$&42.30/36& $ 9.74^{ -0.92}_{+ 0.92}$& $ 8.62^{ -0.81}_{+ 0.82}$\\ \\
3& $ 6.60^{ -0.61}_{+ 0.64}$& $ 1.30^{ -0.05}_{+ 0.04}$& $ >8.70$& $ 2.00^{ -0.04}_{+ 0.08}$& $ 3.00^{ -0.08}_{+ 0.06}$& $ 104.68^{ -14.74}_{+ 21.35}$&33.42/42 & $ 11.44^{ -0.89}_{+ 0.90}$& $ 10.13^{ -0.79}_{+ 0.79}$\\ \\
4& $ 7.49^{ -0.69}_{+ 0.79}$& $ 1.36^{ -0.11}_{+ 0.08}$& $ >7.15$& $ 2.10^{ -0.11}_{+ 0.06}$& $ 2.75^{ -0.12}_{+ 0.12}$& $ 181.65^{ -43.52}_{+ 45.12}$&70.17/53 & $ 15.55^{ -1.02}_{+ 1.02}$& $ 13.76^{ -0.89}_{+ 0.91}$\\ \\
5& $ 6.73^{ -0.42}_{+ 0.48}$& $<1.13$& $ 6.19^{ -0.81}_{+ 0.89}$& $ 2.18^{ -0.07}_{+ 0.10}$& $ 2.79^{ -0.09}_{+ 0.13}$& $ 538.81^{ -95.06}_{+ 173.19}$&93.87/82 & $ 47.81^{ -2.23}_{+ 2.22}$& $ 42.33^{ -1.97}_{+ 1.97}$\\ \\
6& $ 7.22^{ -0.36}_{+ 0.35}$& $ <1.14$& $ 6.19^{ -0.43}_{+ 0.59}$& $ 2.20^{ -0.06}_{+ 0.09}$& $ 2.78^{ -0.07}_{+ 0.07}$& $ 664.90^{ -103.02}_{+ 196.72}$&97.37/90 & $ 58.22^{ -2.55}_{+ 2.55}$& $ 51.55^{ -2.26}_{+ 2.26}$\\ \\
7& $ 6.96^{ -0.45}_{+ 0.51}$& $ 1.17^{ -0.07}_{+ 0.10}$& $ 6.64^{ -0.60}_{+ 0.74}$& $ 2.00^{ -0.04}_{+ 0.11}$& $ 2.94^{ -0.12}_{+ 0.12}$& $ 262.17^{ -30.63}_{+ 71.91}$&101.68/80 & $ 27.80^{ -1.33}_{+ 1.33}$& $ 24.61^{ -1.17}_{+ 1.18}$\\ \\
8& $ 8.26^{ -0.80}_{+ 0.84}$& $ 1.22^{ -0.06}_{+ 0.08}$& $ 7.36^{ -0.82}_{+ 1.01}$& $ 2.05^{ -0.08}_{+ 0.09}$& $ 2.91^{ -0.13}_{+ 0.14}$& $ 308.41^{ -57.81}_{+ 91.09}$&56.15/59& $ 31.38^{ -1.87}_{+ 1.88}$& $ 27.78^{ -1.65}_{+ 1.67}$\\ \\
9& $ 7.54^{ -0.50}_{+ 0.56}$& $ <1.17$& $ 6.43^{ -0.77}_{+ 0.84}$& $ 2.14^{ -0.09}_{+ 0.06}$& $ 2.82^{ -0.09}_{+ 0.13}$& $ 502.07^{ -98.26}_{+ 104.49}$&120.71/89& $ 44.83^{ -2}_{+ 2.01}$& $ 39.70^{ -1.78}_{+ 1.78}$\\ \\
10& $ 7.22^{ -0.49}_{+ 0.55}$& $ <1.31$& $ 6.36^{ -0.64}_{+ 0.93}$& $ 2.19^{ -0.12}_{+ 0.10}$& $ 2.87^{ -0.10}_{+ 0.20}$& $ 381.74^{ -94.94}_{+ 132.76}$&96.26/82& $ 40.29^{ -1.88}_{+ 1.89}$& $ 35.67^{ -1.66}_{+ 1.67}$\\ \\
11& $ 6.69^{ -0.52}_{+ 0.70}$& $ 1.45^{ -0.08}_{+ 0.05}$& $ >8.10$& $ 1.97^{ -0.04}_{+ 0.06}$& $ 3.00^{ -0.10}_{+ 0.09}$& $ 123.20^{ -15.83}_{+ 19.66}$&38.32/47& $ 14.10^{ -1.01}_{+ 1.01}$& $ 12.48^{ -0.89}_{+ 0.90}$\\ \\
12& $ 8.26^{ -1.19}_{+ 1.28}$& $ 1.56^{ -0.07}_{+ 0.07}$& $ >8.22$& $ 1.93^{ -0.04}_{+ 0.03}$& $ 3.00^{ -0.07}_{+ 0.05}$& $ 50.63^{ -10.30}_{+ 7.66}$&33.99/29& $ 5.64^{ -0.68}_{+ 0.68}$& $ 4.99^{ -0.60}_{+ 0.61}$\\ \\
\hline
\end{tabular}
\begin{flushleft}
\small{All the column densities are measured in $10^{22}$ atoms cm$^{-2}$. The Flux reported here is the unabsorbed flux in the energy range 0.7--30 keV. All the fluxes are measured in erg ${cm}^{-2}{s}^{-1}$ and all the luminosities in erg ${s}^{-1}$. The ionisation parameter is measured in erg cm $s^{-1}$.}
\end{flushleft}
\label{reflection}
\end{table*}

\begin{table}
\centering
\caption{Best fit spectral parameters for the model combination \textsc{constant} $\times$ \textsc{smedge} $\times$ \textsc{tbabs} (\textsc{relxillcp} + \textsc{gauss}) for \emph{NuSTAR} data 3--79 keV. The Flux reported here is the unabsorbed flux in the energy range 3--79 keV, measured in $erg {cm}^{-2}{s}^{-1}$ and the luminosity in $erg {s}^{-1}$. The ionisation parameter is measured in erg cm $s^{-1}$. The parameters reported without error means that they are frozen.}
\label{reflection2}
\begin{tabular}{lccccccccc}
\hline

\multicolumn{3}{l}{\textsc{tbabs} $\times$ \textsc{smedge}} \\
\hline
\hspace{0.1cm} N$_{\mathrm{H}}$ & 10$^{22}$ cm$^{-2}$ & $ 5.47^{ -0.18}_{+ 0.17}$ \\
\hspace{0.1cm} E & keV &  $ 7.22^{ -0.13}_{+ 0.12}$\\
\hspace{0.1cm} $\tau$ & &  $ 0.19^{ -0.03}_{+ 0.04}$ \\
\hspace{0.1cm} $\sigma$ & keV &$ <0.25$   \\
\hline
\multicolumn{3}{l}{\textsc{relxillcp}} \\
\hline
\hspace{0.1cm} R$_{\mathrm{in}}$ & ISCO & $ 1.20^{ -0.04}_{+ 0.02}$\\
\hspace{0.1cm} Emissivity index & &$>8.13$ \\
\hspace{0.1cm} $\Gamma$ && $ 1.849^{ -0.003}_{+ 0.006}$ \\
\hspace{0.1cm} log $\xi$ && $ 3.33^{ -0.03}_{+ 0.02}$ \\
\hspace{0.1cm} norm & $\times$10$^{-3}$  & $ 13.91^{ -0.28}_{+ 0.31}$  \\ 

\hline
\multicolumn{3}{l}{\textsc{gauss}} \\
\hline
\hspace{0.1cm} E & keV & $ 6.57^{ -0.01}_{+ 0.02}$ \\
\hspace{0.1cm}  $\sigma$& &$ 0.01$ \\
\hspace{0.1cm} norm & $\times$10$^{-4}$ & $ -8.34^{ -0.99}_{+ 1.05}$  \\
\hline
\hspace{0.1cm} Flux & $\times$10$^{-9}$ &$ 2.55^{ -0.01}_{+ 0.01}$  \\
\hline
\hspace{0.1cm} Luminosity & $\times$10$^{37}$ & $ 2.25^{- 0.01}_{+ 0.01}$  \\
\hline
\hspace{0.1cm} $\chi^{2}$/d.o.f & & 516.94/464 \\
\hline
\end{tabular}
\begin{flushleft}

\end{flushleft}
\end{table}

\section{Discussion} \label{discussion}
We report the first time analysis of a flare during the anomalous low-flux state of GRS 1915+105 using the broad band spectral capability \emph{AstroSat}.  The long coverage, allowed for a time-resolved spectroscopy analysis in order to study the spectral evolution of the flare. The primary result of the analysis is that the spectral fitting are degenerate allowing for two distinct interpretations. In the first interpretation, the source can be modelled as having an absorber which has a large column density and is highly ionised. We find that the flux variation of factor of $\sim 7$, is due to intrinsic variation of the source and cannot be attributed only to variations in the absorber. In the second interpretation, the source  can be modeled as being a relativistically blurred reflection dominated system. In this case, throughout the flare the source remains reflection dominated, indicating that the flare is not primarily due to any geometry changes. To confirm the results obtained from \emph{AstroSat} analysis, we re-analyse a \emph{NuSTAR} spectrum taken during this dim state and show that it can also be fitted by these two models. 

\cite{2020A&A...639A..13K} analysed the \emph{NuSTAR} data used in this study using a model combination similar to what we have used. The difference being that they took into account the gravitational and Doppler redshift of the neutral iron line by allowing the redshift parameter of the reflection component to vary. In our analysis of the \emph{NuSTAR} data, we instead kept redshift fixed at zero and used the single power-law emissivity profile by allowing the emissivity index to vary freely. The rest of the spectral parameters obtained are almost similar to that obtained by \cite{2020A&A...639A..13K} except \emph{relxillCp} norm, which according to our analysis is an order of magnitude higher with roughly the same flux and luminosity.

Two different spectral shapes were observed for GRS 1915+105 by \cite{2020A&A...639A..13K} during the anamolous low-flux state, one before the X-ray/radio flaring with prominent iron absorption line and the other after a high intensity X-ray/radio-flaring period having emission line in the reflection component. Our AstroSat spectrum was similar to the latter one, and the absence of absorption line may be related to the change in ionisation profile of the disc. Similar absorption line (in faint epoch) and absorption edge (in bright epochs) were found in the source by \cite{kong} during June 2019 using \emph{Insight-HXMT} data. Strong edge at $\sim7$ keV along with some weak emission lines were also observed by \cite{neilson} in their interval 1 of a bright flare detected in \emph{NICER} observation of GRS 1915+105 during 20th, May 2019. This absorption edge may be due to the absorption of both the incident and reflected photons by the disc during the Compton scattering process.

For our absorption dominated model, from Table \ref{absorption} it can be seen that the ionisation parameter ($log\xi$) for the ionised absorber and the reflection component is different. Here the reflection component is heavily ionised compared to the ionised absorber so that the reflection may be from the inner parts of the accretion disc or the obscuring medium, and the absorber is in relatively outer parts. Magnetically driven winds with multiple ionisation components can be possible in GRS 1915+105 close to the central engine \citep{ratheesh2021}.  \cite{miller2020obscured} analysed three observations of GRS 1915+105 from 2019 using \emph{Chandra grating}, one on the onset of obscuration and the other two observations deep into the obscured state. They detected a dense, massive accretion disc wind through strong absorption lines during their first observation. They regarded this wind to have two components with distinct properties, originating within a small distance from the BH. From flux-resolved spectroscopy of a bright flare in \emph{NICER} data, \cite{neilson} argued that the obscuring medium is inhomogeneous, has different temperature profiles and is likely radially stratified. They suggested at least three different temperature zones with different ionisation. The results from our absorption dominated model are thus consistent with both \cite{miller2020obscured} and \cite{neilson}, as we also require different ionisation zones for our absorption and reflection components.

Considering the reflection dominated scenario, modelling the spectra with relativistic reflection model \textit{relxillCp} which internally includes the original continuum emission modelled using \textit{nthcomp}, we observed that the inner disc radius is always very close to ISCO, and is positively correlated with 0.7--30 keV unabsorbed flux (Table~\ref{reflection}). The steeper emissivity index observed in all the segments also supports that disc is illuminated mostly in the inner regions. Also as seen in Figure~\ref{fig:ratio_plots} the broad curvature in the Fe k$\alpha$ band requires that the emission is from the innermost regions of the accretion disc. 

The lamp-post model assumes that the hard X-ray emitting region "corona" can be approximated as a point source located above the BH perpendicular to the accretion disc. When the source is in close proximity to the BH, its intrinsic emission is greatly redshifted, and the light-bending effect increases the reflection fraction, resulting in spectra dominated by reflection. Such a lamp-post model is included in \textit{relxillCp} as \textit{relxilllpCp} in which the height of the X-ray emitting source above the BH is included as a spectral parameter. Using model  {\sc constant$\times$tbabs$\times$relxilllpcp} to our \emph{AstroSat} data for all the segments, we found that the height of the X-ray emitting source was always less than $1.98 ^{-1.04}_{+0.62}$ in units of event horizon. Such a value is physically very implausible and it likely reflects the inability of the model to constrain such a parameter as the X-ray emitting source above and below the BH.

\begin{figure}
    \centering
     \includegraphics[scale=0.35]{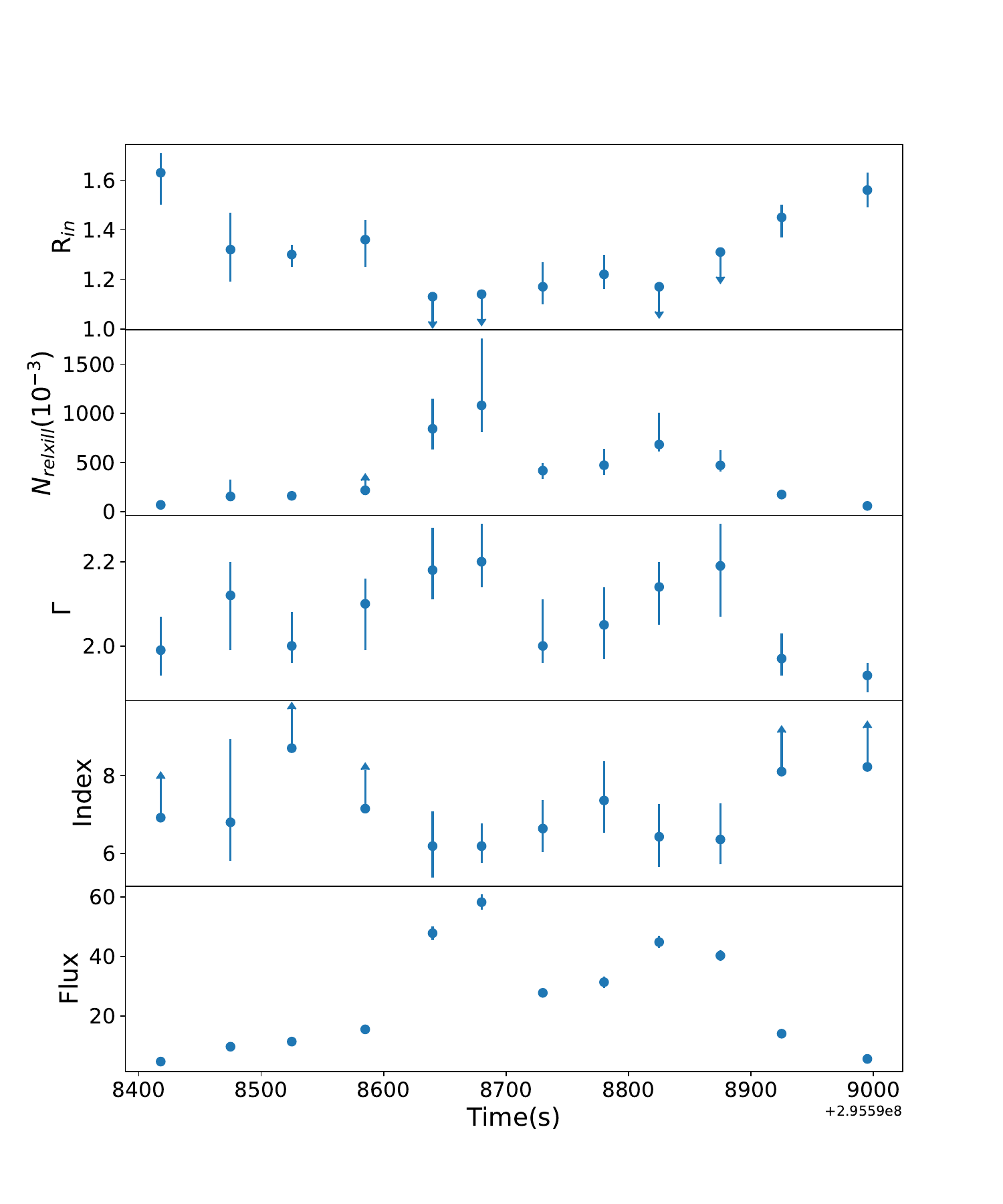}
    \caption{Variation of various spectral parameters as a function of time for the model combination {\sc constant$\times$tbabs$\times$relxillCp} using 0.3--30.0 keV \emph{AstroSat} data. From top to bottom: Panel 1,2,3 ,4 and 5 corresponds to Inner disk radius (ISCO), relxillCp norm, Photon index, emmisivity index of disc and 0.7--30 keV unabsorbed flux ($10^{-9} erg {cm}^{-2}{s}^{-1}$).The flux errors are small and thus not visible in plot.} 
    \label{fig:par_relxill}
\end{figure}

\begin{figure}
    \centering
     \includegraphics[scale=0.64]{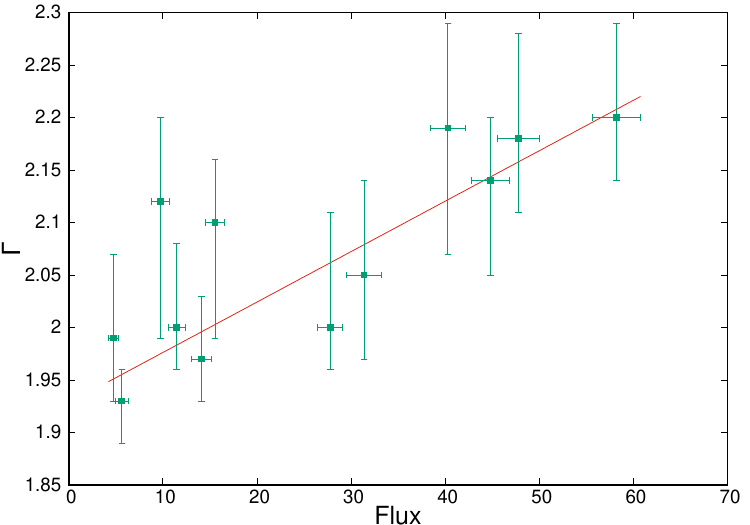}
    \caption{Variation of photon index ($\Gamma$) with 0.7--30 keV unabsorbed flux for the \emph{AstroSat} data.} 
    \label{fig:gamma}
\end{figure}

Using the relativistic reflection model \textit{relxillCp}, we observed that the photon index is positively correlated with flux, as shown in Figure~\ref{fig:gamma}. This behaviour is similar to what has been observed for both reflection-dominated and not reflection-dominated AGN, where the photon index is positively correlated with flux. Also, the softer-when-brighter behaviour revealed mostly by AGN having a reflection fraction less than unity (i.e. not reflection dominated) is consistent in our analysis. The findings presented in this study suggest that for reflection-dominated systems, as the source gets nearer to the black hole (i.e. as the flux decreases), the heating rate of the corona increases, leading to higher temperatures and lower photon indexes. The spectral variability observed here in the case of reflection-dominated spectra can be thus dominated by a variable spectral slope of the primary X-ray continuum. No clear trend was observed for the ionisation parameter $log\xi$ and it was varying between \textit{2.67 and 3.09} for all the segments of \emph{AstroSat} data, for the \emph{NuSTAR} data its value found was $ 3.33^{ -0.03}_{+ 0.02}$, larger than that of \emph{AstroSat} data which suggests that it decreases with increase in flux. We observed a positive correlation between \textit{relxillCp} norm and flux as evident from Table~\ref{reflection} and \ref{reflection2} (see Figure \ref{fig:par_relxill}).

\begin{figure}
    \centering
     \includegraphics[scale=0.35]{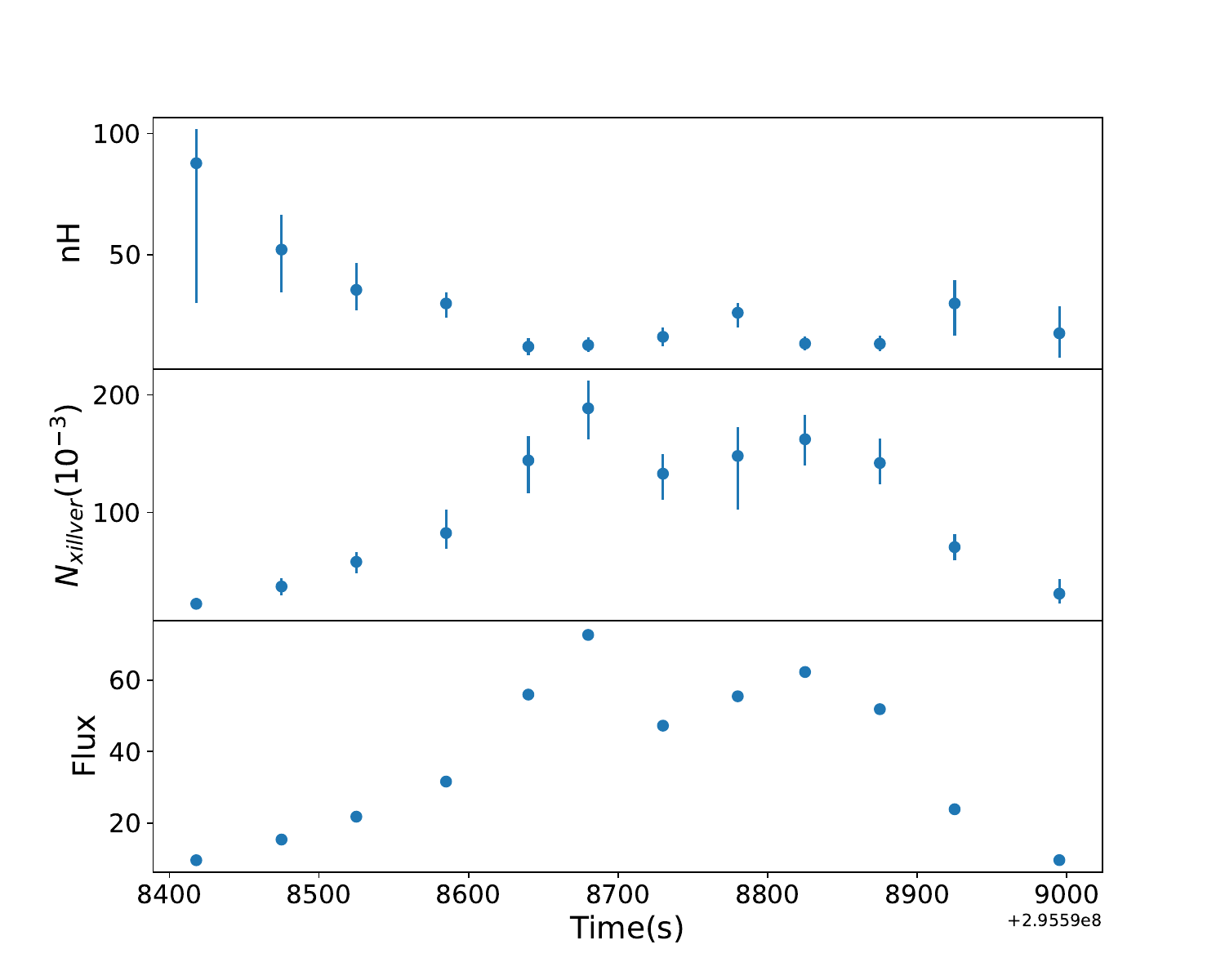}
    \caption{Variation of various spectral parameters as a function of time for the model combination {\sc constant$\times$tbabs$\times$zxipcf(thcomp$\times$diskbb + xillverCp)} using 0.3--30.0 keV \emph{AstroSat} data. From top to bottom: Panel 1,2 and 3 corresponds to nH of the ionised absorber (10$^{22}$ cm$^{-2}$), xillverCp norm and 0.7--30 keV unabsorbed flux ($10^{-9} erg {cm}^{-2}{s}^{-1}$). The flux errors are small and thus not visible in plot. Note that although there is variation of the absorption column density, the large variation of the unabsorbed flux indicates an intrinsic origin for the flare.} 
    \label{fig:par_zxipcf}
\end{figure}

Considering the absorption-dominated scenario, we observed that the ionisation parameter $log\xi$ for \textit{zxipcf} the partial coverer ionised absorber was less compared to that of the reflection component \textit{xillverCp} used to fit only the reflection component. Here we observed a highly ionised medium in the case of the reflection component, so its origin can be different from the ionised absorber. These results are consistent with results obtained by \cite{2021A&A...655A..96R}, while analysing the similar low flux state \emph{NuSTAR} data of GRS 1915+105 where they found that at all flux levels, the obscuring matter is highly ionised. We found that the ionisation parameter for the ionised absorber varies between 1.97 and 3.14, with no clear trend of change with respect to flux, while as the hydrogen column density of the ionised absorber is anti-correlated with flux. We also observed a positive correlation between the normalisation of the reflection component \textit{xillverCp} and flux as evident from Table~\ref{absorption} (see Figure \ref{fig:par_zxipcf}). 

\begin{figure}
    \centering
     \includegraphics[scale=0.64]{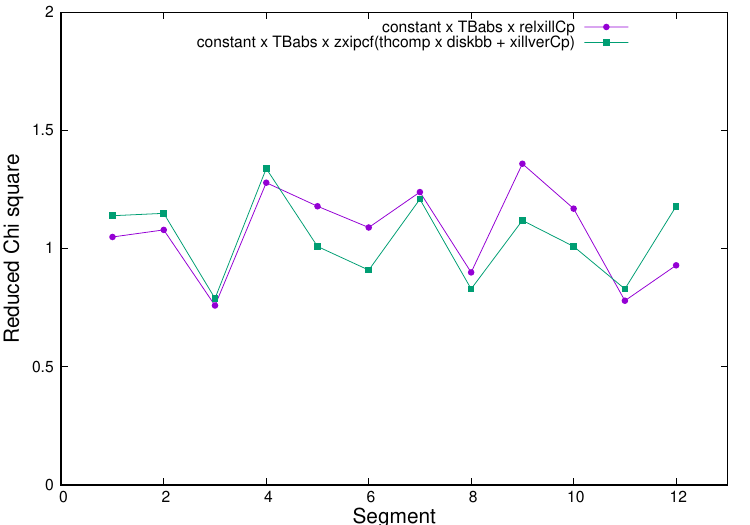}
    \caption{Comparison of reduced Chi Square using different models.}
    \label{fig:chi_sq}
\end{figure}

We also observed that the flux calculated using absorption dominated model was greater than that of reflection dominated model. The reduced chi-square obtained during the fitting of data with reflection dominated model and ionised absorption model is plotted in Figure~\ref{fig:chi_sq}, and we can see from the figure that both the models give similar statistical fits to the data. We have also verified this similar fit using the \emph{NuSTAR} data, with parameters and the reduced chi-square obtained using both the models, presented in Table~\ref{reflection2} and Table~\ref{absorption2}. The reflection dominated interpretation is fairly similar to the reflection dominated spectrum used to model the low flux state of AGN. However, the spectra of GRS 1915+105 can also be interpreted as being due to high ionised  absorption. Thus, unless different physical mechanisms are invoked for the similar spectra observed, it seems that either both type of sources in their low state, are either reflection dominated or highly absorbed. A comprehensive analysis using wide band data of both AGN and XRBs will reveal more information regarding the nature of these sources in their extreme low flux states.

\section{Conclusions} \label{Conclusions}
We performed time resolved spectroscopy of a bright flare observed just a day after \emph{MAXI/GSC} reported the re-brightening phase, during the anamolous low-flux state of GRS 1915+105, using broadband spectral capabilities of \emph{AstroSat}. Our spectral fitting allowed for two different interpretations. In one interpretation source can be modelled as being relativistically blurred reflection dominated system and in the other interpretation having highly ionised absorber with high column densities. The two different interpretations of the spectra were confirmed by re-analysing a similar \emph{NuSTAR} observation during the dim state of the source. For the absorption interpretation, the intrinsic flux of the source varied by a factor of $\sim 7$ and hence the flare cannot be ascribed to a dramatic change in obscuration. Similarly for the reflection dominated model, the flare is due to intrinsic flux variation and the source remains reflection dominated throughout, indicating that the flare is not primarily due to any geometry changes.

We note that the relativistically blurred reflection dominated spectrum for a dim state of GRS 1915+105 is analogous to results obtained for AGN and hence the
results presented here will have a bearing on AGN studies. Relativistically blurred reflection dominated spectra characterised by high reflection fraction has been observed in low flux state of many AGN (Seyferts) MCG-6-30-15 , 1H 0707-495 \citep{2004MNRAS.353.1071F} and 1H0419-577 \citep{fabian2005x}. This enhanced reflection fraction in AGN is seen as a consequence of X-ray emitting source being very close to the central BH, and thus the light bending effects come into play. The low lamp-post heights and high reflection fraction in the spectra analyzed in this work, makes one of our spectral interpretation similar to the one used for the low flux state of AGN \citep{waddell2019,barua}. However, the spectra of GRS 1915+105 can also be interpreted in terms of high complex absorption and thus it may be that the low flux AGN spectra have a similar interpretation and more detailed studies are required.

\section*{Acknowledgements}
We thank the anonymous reviewer for valuable and insightful comments. This publication uses the data from the \emph{AstroSat} mission of the Indian Space Research Organisation (ISRO), archived at the Indian Space Science Data Centre (ISSDC). Data from \emph{LAXPC} and \emph{SXT} onboard \emph{AstroSat} mission was used. We are thankful to \emph{LAXPC} and \emph{SXT POC} teams at TIFR Mumbai for providing the necessary software tools required for the analysis. SB is grateful to IUCAA Pune for regular visits under its visitor programme, where a part of this work was done. This research has made use of data and/or software provided by the High Energy Astrophysics Science Archive Research Center (HEASARC), which is a service of the Astrophysics Science Division at NASA/GSFC. SB thanks UGC, Govt. of India for providing fellowship under the UGC-JRF scheme(Ref. No.: 191620095661/CSIR-UGC NET DECEMBER 2019). VJ would like to acknowledge the Centre for Research Projects, CHRIST (Deemed to be University), for the financial support in the form of a Seed Money Grant (SMSS-2217).

\section*{Data Availability}
The data utilized in this article are available at \emph{AstroSat−ISSDC}
website (\url{http://astrobrowse.issdc.gov.in/astro archive/archive/Hom
e.jsp}) and \emph{NuSTAR} Archive (\url{https://heasarc.gsfc.nasa.gov/docs/nustar/
nustar archive.html}). The software used for data analysis
is available at HEASARC website (\url{https://heasarc.gsfc.nasa.gov/lh
easoft/download.html}).

\bibliographystyle{mnras}
\bibliography{references}
\bsp	
\label{lastpage}
\end{document}